\documentclass[preprint,prb,aps,floatfix,amsmath,amssymb,superscriptaddress]{revtex4}

\usepackage{mathtools}
\usepackage{amsfonts}
\usepackage{amssymb}

\usepackage{graphicx}
\usepackage[all]{xy}

\usepackage{amsthm}

\newtheorem{theorem}{Theorem}[section]
\newtheorem{lemma}[theorem]{Lemma}
\newtheorem{definition}[theorem]{Definition}
\newtheorem{conjecture}[theorem]{Conjecture}
\newtheorem{openquestion}[theorem]{Open Question}
\newtheorem{remark}[theorem]{Remark}

\newcommand{\zz}{{Z}_2}

\newcommand{\zc}{{Z}_p}
\newcommand{\fc}{F_p}
\newcommand{\coeff}{p}

\newcommand{\be}{\begin{equation}}
\newcommand{\ee}{\end{equation}}

\begin{document}
\title{Quantum Systems on Non-$k$-Hyperfinite Complexes: A Generalization of Classical Statistical Mechanics on Expander Graphs}

\author{M. H. Freedman}
\affiliation{Microsoft Research, Station Q, CNSI Building, University of California, Santa Barbara, CA, 93106}

\author{M. B. Hastings}
\affiliation{Microsoft Research, Station Q, CNSI Building, University of California, Santa Barbara, CA, 93106, mahastin@microsoft.com}

\begin{abstract}
We construct families of cell complexes that generalize expander graphs.  These families are called non-$k$-hyperfinite, generalizing the idea of a non-hyperfinite (NH) family of graphs.  Roughly speaking, such a complex has the property that one cannot remove a small fraction of points and be left with an object that looks $k-1$-dimensional at large scales.  We then consider certain quantum systems on these complexes.  A future goal is to construct a family of Hamiltonians such that every low energy state has topological order as part of an attempt to prove the quantum PCP conjecture.  This goal is approached by constructing a toric code Hamiltonian with the property that every low energy state without vertex defects has topological order, a property that would not hold for any local system in any lattice $Z^d$ or indeed on any $1$-hyperfinite complex.  Further, such NH complexes find application in quantum coding theory.  The hypergraph product codes\cite{hpc} of Tillich and Z\'{e}mor are generalized using NH complexes.
\end{abstract}
\maketitle
In this paper, we construct several families of cell complexes which generalize expander graphs.  More precisely, these families are a  generalization of families of graphs which are ``not hyperfinite" as defined by Elek\cite{hyperfinite}; being not hyperfinite is a property that is closely related to being an expander graphs but slightly weaker.
We then study various quantum Hamiltonians on these complexes and find that they realize properties that rely on the fact that these complexes are not $k$-hyperfinite, as we define below in definition \ref{defn1}.

The motivation for this work is that expander graphs play a key role in coding theory and complexity theory and we are looking for a similar construction which could play a similar role in quantum coding theory and quantum complexity theory.
Later in the paper we discuss applications to coding theory, but now we briefly discuss the application to complexity theory, in particular the quantum PCP conjecture\cite{qpcp}.
The quantum PCP (qPCP) conjecture deals with the difficulty of approximating the ground state energy of quantum many-body Hamiltonians.  Suppose we have a system with $N$ degrees of freedom, each degree of freedom having a Hilbert space of dimension $O(1)$ with the Hilbert space of the whole system being
the tensor product of these Hilbert spaces.  The Hamiltonian $H$ is a sum of interaction terms, each interaction term being supported on $O(1)$ sites with each site having $O(1)$ interaction terms supported on that site, and each interaction term having operator norm bounded by $O(1)$ so that the operator norm of $H$ is
$O(N)$.  The qPCP conjecture is that there is some constant $c$ (this constant may depend upon the particular quantities referred to as being ``$O(1)$" in the previous sentence) such that it is QMA-hard to approximate the ground state energy to an accuracy better than $cN$.
In Ref.~\onlinecite{qpcpme}, it was noted that in order for the qPCP conjecture to be true, there must exist a family of Hamiltonians $H$, obeying these requirements on the support and norm of the interaction terms, such that  there is no trivial state $\psi$ (as defined below) such that $\langle \psi,H \psi \rangle$ is at most $cN$ above the ground state energy.  The reason that this needs to hold is that otherwise the trivial state can serve as a classical witness for the existence of a low energy state, putting the approximation problem wthin NP.
  Defining the ``energy density" of a state to be $1/N$ multiplied by the difference between the energy of that state and the ground state energy, the requirement is that there exist no state with energy density less than $c$ for some $c$.
In this paper, we consider the easier problem of constructing a family of Hamiltonians with no low energy trivial states (NLTS).  The NLTS problem, defined in definition \ref{nlts}, is a natural first step toward provng qPCP, and is in our view not a less interesting objective.
The Hamiltonians that we construct in fact will have a ground state energy that we can compute exactly, so this does not yet address more difficult problem of constructing a family of Hamiltonians such that approximating the ground state energy is QMA-hard.  Our constructions use only Hamiltonians which are a sum of commuting terms.
In fact, we do not fully succeed in constructing such a family of Hamiltonians but we obtain partial results which rely explicitly on the complex being not $1$-hyperfinite.  In particular we construct a complex on which a toric code has a certain ``one-sided" NLTS, in that there is no trivial state with no vertex defects and with low energy for the plaquette terms, where throughout this paper we use ``low energy" to mean an energy density bounded by some small constant; in contrast, Ref.~\onlinecite{qpcpme} shows that on a $1$-hyperfinite complex the toric code has a trivial ground state with no defects.
We conjecture a construction of a system with NLTS (removing this ``one-sided" restriction) but we do not prove this.

This application to quantum systems led to our definition \ref{defn1}, which is different from other generalizations of expander graphs considered elsewhere\cite{otherexpanders1,otherexpanders2}.
The outline of the paper is as follows.  In section \ref{trivstate} we define trivial states and quantum circuits.  In section \ref{defines}, we give the definition of $k$-hyperfiniteness and for every $k$ we construct families of complexes which are not $k$-hyperfinite.
In section \ref{QSqCl}, we further expand on the motivation for considering these complexes, by giving an analogy to classical systems.  Sections \ref{defines},\ref{QSqCl} can be read in either order; more mathematical readers might prefer our order, while more physical readers might prefer to read section \ref{QSqCl} first.  Section \ref{QSqCl} can be skipped if desired as it is purely motivational, however it does present some results about statistical mechanics on expander graphs that may be of interest; while these results in this section may be well-known, we have not found them in the literature.  In section \ref{tconcomplex} we review the definition of the two-dimensional toric code Hamiltonian, and give this definition for arbitrary $2$-complexes.  Further, we describe other toric codes, so-called $(q,r)$-codes in dimension $d=q+r$ (often, these would be referred to as $(p,q)$ codes, but we are using the symbol $p$ for other purposes here).
In section \ref{toward}, we consider the toric code on a family of complexes that is not $1$-hyperfinite.  While we do not show that the code on this family has NLTS, we show that there is no trivial state for which the terms in the Hamiltonian that penalize broken strings are in their ground state and which is low energy with respect to the other terms (the terms that give a dynamics to the string).  As we note, this property would not occur for any complex which was $1$-hyperfinite including such familiar examples as a two-dimensional lattice or an expander graph.  In section \ref{hypgc} we discuss the relation to hypergraph product codes and give an alternative presentation of those codes in terms of the language of chain complexes; this presentation leads us to a generalization of these codes.  We also present a possible conjecture on a Hamiltonian with NLTS.
Finally, in the discussion we discuss an interpretation of our results in terms of dynamics of strings or higher dimensional surfaces, and give a conjecture that if a certain family of non-$k$-hyperfinite complexes could be constructed then one could indeed give a family of complexes on which a toric code had NLTS, now requiring only that the state be low energy with respect to all terms in the Hamiltonian, rather than requiring that it also be zero energy with respect to certain terms.  In an appendix, we discuss possible generalizations of our definition of hyperfiniteness to the case of noncompact spaces where we define a notion of $k$-amenability.

\section{Definitions and Trivial States}
We define some metric ${\rm dist}(i,j)$ between degrees of freedom $i,j$.  We choose this metric so that the interaction terms in the Hamiltonian have support on sets of diameter $O(1)$.  One way to define this metric would be to define an interaction graph, with vertices corresponding to degrees of freedom and an edge connecting two vertices if both degrees of freedom were in the support of some term in the Hamiltonian, and then to use the metric which is the shortest path metric on this graph.  In fact, later in the paper we will not use this metric but a different one; the reason is that for much of the paper, the degrees of freedom will be on edges, rather than on vertices.  So, for that situation we will choose a metric in which two edges attached to the same vertex are a distance $1$ from each other.  However, the specific metric used is not of much importance (the mathematical notion of quasi-isometry expresses in detail which metrical dfferences are ``unimportant" for us).

\label{trivstate}
We begin by defining a quantum circuit.
\begin{definition}
A unitary quantum circuit $U$ is a unitary of the form
\be
U=U_1 U_2 .... U_k,
\ee
where $k$ is called the ``depth" of the circuit, such that each unitary $U_i$ can be written as a product
\be
\label{roundDefn}
U_i=U_{i,1} U_{i,2} ... U_{i,n_i}
\ee
where the support of $U_{i,a}$ is disjoint from the support of $U_{i,b}$ for $a\neq b$ for all $i$ and where the integer $n_i$ is the number of unitaries in the $i$-th round.
The ``range" of a unitary $U_{i,a}$ is the diameter of the support of that unitary.
We define the {\bf range} of a quantum circuit to be the product of the depth of the quantum circuit times the maximum range of the unitaries in that quantum circuit.
\end{definition}
Note that this definition of the range of a quantum circuit is not completely standardized since some authors take the range of the circuit to be the maximum range of a unitary in the circuit; we prefer this definition here since it makes the definition of trivial states more natural.

We define a trivial state as follows:
\begin{definition}
A state $\psi$ with $|\psi|=1$ is $(R,\epsilon)$-trivial if there exists a unitary quantum circuit $U$ with range $R$ such that
\be
|\psi-U\psi_{prod}| \leq \epsilon
\ee
for some product state $\psi_{prod}$.

We say that $\psi$ is $R$-trivial if it is $(R,0)$-trivial.
\end{definition}

One may allow the use of ancillas in these constructions.   For each degree of freedom in the original system, we define two Hilbert spaces, one called the real Hilbert space which is the same as the Hilbert space for that degree of freedom in the original system and one called the ancilla Hilbert space.  We can then consider trivial states on this larger Hilbert space obtained by unitary quantum circuits acting on product states on the larger Hilbert space.  The ancillas are organized by the same geometry as the original Hilbert space; that is, the distance ${\rm dist}(...)$ between an ancilla on a given site and the real degree of freedom on that site is equal to zero.
By tracing out the ancilla degrees of freedom, the result may be a mixed state; we refer to this as a trivial mixed state, and indeed any convex combination of such states will be referred to as a trivial mixed state following Ref.~\onlinecite{mixedtriv}.  Our results later are phrased in terms of trivial pure states, but the same results can be proven for trivial mixed states (add ancillas to prove it for a trivial mixed state obtained by tracing out ancillas and then once we have proven that no such trivial state has low energy it will also follow that no convex combination of such states will have low energy).

With these definitions, we are now prepared to define the NLTS problem.
\begin{definition}
\label{nlts}
Consider a family of Hamiltonians, $H_N$, with $N\rightarrow \infty$.   Each $H_N$ acts on a Hilbert space which is a tensor product of $N$ different finite dimensional Hilbert spaces, called sites, each having dimension $O(1)$.
Each $H_N$ is a sum of interaction terms, each interaction term being supported on $O(1)$ sites with each site having $O(1)$ interaction terms supported on that site, and each interaction term having operator norm bounded by $O(1)$.
Assume that the smallest eigenvalue of $H_N$ is equal to $0$ for all $N$.

We say that such a family of Hamiltonians $H_N$ has {\bf NLTS} if there exists a constant $\epsilon>0$ such that there is no $R$ such that for all $N$ there is an $R$-trivial state $\psi_N$ such that
\be
\langle \psi_N, H_N \psi_N \rangle \leq \epsilon N.
\ee
\end{definition}
We conjecture
\begin{conjecture}
A family of Hamiltonians with NLTS does exist.
\end{conjecture}
We have not specified that $H_N$ should be a sum of commuting terms in the above definition, or that $H_N$ should be frustration free, although the Hamiltonians that we consider n the rest of the paper are indeed frustration-free and sums of commuting terms.  One might naturally defining problems such as commuting NLTS where such a commuting restriction is imposed.

Our results later will show the absence of $R$-trivial states with certain low energy properties.  However, from the point of view of the qPCP conjecture and NLTS problem, these definitions of $(R,0)$-trivial and $(R,\epsilon)$-trivial are not too different.  If there is an $(R,\epsilon)$-trivial state $\psi$, then
\be
|\langle U \psi_{prod},H U \psi_{prod}\rangle - \langle \psi,H \psi \rangle| \leq (2 \epsilon+\epsilon^2) \Vert H \Vert,
\ee
and so, since $\Vert H \Vert/N=O(1)$, if there is no $R$-trivial state for a Hamiltonian with energy density less than $c$, then there is no $(R,\epsilon)$ trivial state for that
Hamiltonian with energy density less than $c+O(1)(2\epsilon+\epsilon^2)$.
This point may not apply to our one-sided results later, since they are based on states with zero vertex defects (hence zero, rather than small, energy for a certain type of term), but still we find the case of $R$-trivial states worth considering.

We note that the reason that trivial states are relevant for the qPCP conjecture is that a trivial state can serve as a witness because one can efficiently compute the reduced density matrix of the trivial state on small sets.  However, NLTS is weaker than qPCP because there are lots of other possible choices for a classical witness.  For example, one could take a state that is trivial using a {\it different} metric to define the support of the unitaries $U_{i,a}$ in Eq.~(\ref{roundDefn}), different from that used to define the interactions.  We will not consider such states in this paper.

Another issue relevant to qPCP is the particular choice in our definition of NLTS to require that the range $R$ be independent of $N$.  In fact a classical witness given as an $R$-trivial state allows one to efficiently evaluate the expecation value of the Hamiltonian for a sufficiently slowing growing $R$.
Let us estimate the allowed growth of $R$.  Each term of the Hamiltonian is supported on some small number of sites; if the number of sites within range $R$ of this set is at most logarithmic in $N$ then the dimension of the Hilbert space on that set of sites is at most polynomial in $N$ and the expectation value of that term can be efficiently evaluated by direct simulation of the quantum circuit restricted to that set of sites.  For systems defined on expander graphs, as we consider later, this requirement of having only logarithmically many sites means that we can in fact efficiently evaluate expectation values for $R$ which is bounded by a sufficiently small constant times $\log(\log(N))$, where the constant depends upon the expansion ratio; of course, we have no proof that there is no way to efficiently evaluate expectation values for $R$ growing more rapidly than this but we have no algorithm to do it either.  In contrast, for a system on a regular two-dimensional lattice, we can efficiently evaluate expectation values for an $R$ growing as $\sqrt{\log(N)}$ on a classical computer\cite{mtriverror}.

\section{Families of Complexes That Are Not $k$-Hyperfinite}
\label{defines}
In this paper we will construct various complexes.  In Ref.~\onlinecite{qpcpme}, the complexes constructed were simplicial complexes.  In this paper, it is also convenient to use cubical and more generally polyhedral complexes where convex polytopes are glued together via simplicial homeomorphisms of their faces.
Another distinction is that in Ref.~\onlinecite{qpcpme}, a complex called an interaction complex was constructed: the degrees of freedom of the quantum Hamiltonian were placed on the $0$-cells of the complex, and the higher cells of the complex were determined by the support of the terms in the Hamiltonian.  In this paper, we will often place the degrees of freedom on $1$-cells or even high cells of the complex.

A graph can be viewed as a $1$-complex.  We will use the terms ``vertex" and ``$0$-cell" interchangeably; we will similarly use the terms ``edge" and ``$1$-cell" interchangeably and also the terms ``plaquette" and ``$2$-cell" interchangeably.
We will later refer to ``vertex operators" and ``plaquette operators" for a toric code, which will be operators defined for each $0$-cell or $2$-cell of the complex, respectively.
One case in which we will try to distinguish between these terms is that for certain complexes which are constructed from a product of graphs, we will use the term ``$1$-cells" for the complex and the term ``edge" for the graph.

We place a metric on a $1$-complex, by defining each edge to have length $1$, and using a path metric.  This reproduces the usual graph metric between vertices on the graph (the distance between neighboring vertices is equal to $1$).  We extend this metric to some path metric over the higher cells so that the cells in each dimension have bounded diameter.
Now define $k$-localizable complexes by:
\begin{definition}
A metrized simplicial $l$-complex $K_l$ is  ``$k$-localizable with range $R$" if there exists a continuous function $f$ from $K_l$ to some simplicial $k$ complex $K_k$ (using the same graph metric as above on the edges of $K_l$, extended as above to the higher cells of $K_l$) such that the diameter of the pre-image of any point in $K_k$ is less than $R$.
\end{definition}
Next define $k$-hyperfiniteness:
\begin{definition}
\label{defn1}
Consider a family of $l$-complexes, $K_l(N)$, where $N$ is the number of $0$-cells in $K_l(N)$, with $N \rightarrow \infty$.  Such a family is said to be ``$k$-hyperfinite" if
for all $\epsilon>0$, there exists an $R$ such that for all $N$ one can remove at most a fraction
 $\epsilon$ of the $0$-cells of $K_l(N)$, while removing also all attached higher cells, such that the resulting complex is $k$-localizable with range $R$.
\end{definition}
We note that the definition of $k$-localizable was previously called ``intermediate diameter" in Ref.~\onlinecite{Gromov}; we stick with the terminology above to coincide with the usage in Ref.~\onlinecite{qpcpme}.

Families of interest to us generally have bounded local geometry: the number of $j+1$ cells meeting any $j$-cell and the number of $j$ cells in the boundary of any $j+1$-cell is upper bounded independent of $N$ (and $j$).  But we do not include this property in the definition.

The definition of a family of $1$-complexes being $0$-hyperfinite coincides with the definition of a family of graphs being hyperfinite\cite{hyperfinite}, motivating our terminology.

Note that if the metrics $d_N$ on the family $K_l(N)$ are altered by a uniformly bounded amount: $d_N$ replaced with a quasi-isometric $d_N'$ so that for all $x$, $y$ $\in K_l(N)$, $\frac{1}{C_0}d_N(x,y)-C_1 < d_N'(x,y)<C_0d_N(x,y)+C_1$ for $C_0,C_1>0$, the property of being or not being $k$-hyperfinite is unchanged, so the choice of metric is quite flexible.

We now construct families of complexes which are not $l$-hyperfinite.  To exposite key ideas first we begin with the case $l=1$.
Consider a family of expander graphs with fixed degree $d$ and with girth $g$ of the graph diverging as $N\rightarrow \infty$; in fact, we may take the girth of order $\log(N)$.  A graph $G$ in this family with $N$ vertices has $\frac{d}{2} N$ edges and has first Betti number $b_1(G)=\frac{d-2}{2} N + 1$.  We will define a $2$-complex, by taking the Cartesian product of this graph $G$ with itself, $G\times G$.  This is the ``product graph'' with $2$-cells, i.e. Euclidean squares, of the form edge $\times$ edge glued in.
Note that a $0$-cell in $G \times G$ is labeled by a pair $(i,j)$, where $i,j$ are vertices in $G$.  A $1$-cell in $G\times G$ is labeled by either a pair $(e,i)$, where $e$ is an edge in $G$ and $i$ is a vertex, or a pair $(i,e)$.
Since the girth of the graph $G$ is diverging, we may assume that it is larger than $4$ for large enough $N$, so all length $4$ loops are boundaries of such squares.  Explicitly, let an edge $e$ connect vertices $e_1,e_2$ in $G$ and let an edge $f$ connect vertices $f_1,f_2$.  Then, the length $4$ loops are of the form $(e,f_1),(e_1,f),(e,f_2),(f,e_2)$.
An important property of $G \times G$ is that it has $N^2$ $0$-cells and has an ``extensive" second Betti number, where a quantity will be called ``extensive" if it is $O(\text{the number of }0\text{-cells})$.  In this case,
\be
b_2(G \times G)=\Bigl( \frac{d-2}{2} N + 1\Bigr)^2.
\ee
It is this large second Betti number, combined with the high girth of $G$, that we use to prove that the family of complexes $G \times G$ is not $1$-hyperfinite.

Before giving the detailed proof of the theorem below, let us sketch the overall strategy.  First, we use the fact that the second Betti number of $G \times G$ is extensive to show that deleting a sufficiently small fraction of cells from $G \times G$ leaves a complex $G \times G \setminus C$ that still has a nonzero second Betti number since deleting a $2$-cell changes the Betti number by at most one.  This implies that the resulting complex still contains a nontrivial $2$-cycle.  The assumption that $G \times G \setminus C$ is $1$-localizable means that there is a map $f$ from $G \times G \setminus C$ to a $1$-complex $K$.  This map $f$ maps the nontrivial $2$-cycle to a subset of $K$.  It is at this point that we use the large girth assumption on $G$ to show a contradiction: since $G \times G$ has large girth, the nontrivial $2$-cycle in $G \times G \setminus C$ cannot be continuously mapped to a $1$-complex using a map with bounded diameter of pre-images.
To motivate why the large girth is needed, consider the following case of a complex with a small girth: a``long, thin" torus (a torus with the second radius being small; this torus is the product of a large girth line graph with a small girth line graph).  One can find a continuous map from this torus to a circle in the natural way, mapping a point on the torus with angles $(\theta,\phi)$ to a point on the circle at angle $\theta$, giving a map with a diameter of pre-image bounded by the girth of the second graph.  In contrast, since we are considering complexes $G \times G$ with large girth, we will show that a continuous map to a $1$-complex with bounded diameter of pre-images is impossible, showing that such a map would lead to a contradiction as it would instead imply that $G \times G \setminus C$ had vanishing second Betti number.

A second point worth recalling is the idea of a simplicial map and the simplicial approximation theorem.  These ideas enable us to avoid pathologies that might occur for arbitrary continuous functions and let us instead deal with more combinatoric ideas.
 Simplicial maps are maps such that the images of the vertices of a simplex span a simplex.  Simplicial maps are completely determined by their action on the $0$-cells.  The simplicial approximation theorem enables us to approximate a continuous map between two complexes by a simplicial map, after some subdivision of the simplices of both complexes.  The approximation can be made arbitrarily accurate by taking sufficiently fine subdivision and the approximating map is homotopic to the original map.

\begin{theorem}
\label{thm2.1}
Let $G_N$ be a family of connected degree $=d$ graphs with diverging girth.  The family of $2$-complexes $G_N\times G_N$ is not $1$-hyperfinite.
\begin{proof}
To a connected graph $G$ is associated a $Z_2$-cellular chain complex:
$$C_1\xrightarrow{\partial}C_0\xrightarrow{a}Z_2\xrightarrow{}0$$
where $C_i$ is the $Z_2$-vector space generated by $i$-cells: $C_1\cong Z_2^{\frac{d}{2}N}$, $C_0\cong Z_2^N$, and the final map (epimorphism) to $Z_2$ is the ``augmentation'' $a$ which records connectivity of the graph by taking every generator to $1$.  The map $\partial$ is onto $\text{ker}(a)$, so $b_1=\text{rank}(H_1(G;Z_2))=\text{rank ker}(\partial)=\frac{d-2}{2}N+1$.  By the K\"{u}nneth formula, $b_2(G\times G)=\text{rank }H_2(G\times G;Z_2)=(\frac{d-2}{2}N+1)^2$.

Deleting a $0$-cell implies deleting $d^2$ adjacent squares ($2$-cells) of $G\times G$.  Since $d\geq 3$ is fixed, for $N$ large and $\epsilon<0.02$ which we now assume, deleting an $\epsilon$-fraction of $0$-cells implies deleting at most $d^2\epsilon N^2<(\frac{d-2}{2}N+1)^2$ $2$-cells from $G\times G$.

An easy exercise in Mayer-Vietoris (M-V) sequences establishes:

\begin{lemma}
$b_2(G\times G \setminus C)\geq(\frac{d-2}{2}N+1)^2-|C|>(\frac{d-2}{2}N+1)^2-d^2\epsilon N^2>0$, where $C=C_1\cup C_2\cup\cdots\cup C_{|C|}$ is the union of the $2$-cells being deleted and $|C|$ is the number of such $2$-cells.  Clearly for $\epsilon$ small enough $b_2(G\times G \setminus C)$ becomes extensive.

\begin{proof}
If $A$ and $B$ are sub-complexes of a complex $X$ the M-V exact sequence reads
$$\longrightarrow H_i(A\cap B)\longrightarrow H_i(A)\oplus H_i(B)\longrightarrow H_i(X)\xrightarrow{\hspace{0.075in}\partial\hspace{0.075in}}H_{i-1}(A\cup B)\longrightarrow$$
where $i$ is arbitrary and we take all coefficients $=Z_2$.  Let us set $i=2$ and inductively assume $X=G\times G \setminus C_1\cup\cdots\cup C_{k-1}$, $A=C_k$, and $B=G\times G \setminus C_1\cup\cdots\cup C_k$.  Now the right hand term $H_1(A\cap B)$ is the first homology of some subset of the circle $\partial C_k$ which is $Z_2$ if the subset is the entire circle and $0$ otherwise.  $H_2(C_k;Z_2)\cong 0$, a disk being contractible has no homology except in dimension 0.  So our sequence reads:
$$\longrightarrow H_2(G\times G \setminus C_1\cup\cdots\cup C_k;Z_2)\longrightarrow H_2(G\times G \setminus C_1\cup\cdots\cup C_{k-1};Z_2)\longrightarrow Z_2\text{ or }0\longrightarrow$$

By exactness, the rank of the left term, $b_2(G\times G \setminus C_1\cup\cdots\cup C_k;Z_2)$, plus the rank of the right term must be greater than or equal $\text{rank}(H_2(G\times G \setminus C_1\cup\cdots\cup C_{k-1};Z_2))=b_2(G\times G \setminus C_1\cup\cdots\cup C_{k-1};Z_2)$.  Thus the $Z_2$ second Betti number $b_2$ will either decrease by one or remain unchanged whenever a $2$-cell $C_k$ is deleted.  The lemma follows.
\end{proof}
\end{lemma}

\begin{lemma}
\label{lem2.3}
The map $H_2(G\times G \setminus C;Z_2)\xrightarrow{\operatorname{inc}_2}H_2(G\times G;Z_2)$ is an injection.
\begin{proof}
The map $\operatorname{inc}_2$ fits into the exact sequence of a pair:
\newpage
$$\longrightarrow H_3(G\times G,G\times G \setminus C;Z_2)\xrightarrow{\hspace{0.075in}\partial\hspace{0.075in}}H_2(G\times G \setminus C;Z_2)\xrightarrow{\operatorname{inc}_2} H_2(G\times G;Z_2)\longrightarrow$$
\hspace{1.5in}$\|{\large{\wr}}\text{ excision}$

\hspace{1in}$H_3(C,\partial C;Z_2)$

\hspace{1.5in}$\|{\large{\wr}}$

\hspace{1.5in}$0$

Via excision, the first term is $H_3$ of a relative $2$-complex and therefore vanishes, showing $\operatorname{inc}_2$ to be an injection.
\end{proof}
\end{lemma}

\begin{remark}
\label{rem2.4}
The vanishing of homology in dimensions exceeding the dimension of the complex is a property of all \emph{ordinary} homology theories, i.e. theories obeying the dimension axiom.  These include $H_*(\text{ };Z_2)$ which we are using but not, for example, theories such as $K$-homology, $K_*(\text{ })$.
\vspace{.1in}
\end{remark}

\noindent\emph{Proof of theorem.}
Pick $\epsilon>0$ small enough so that for $N$ large and for any set $C$ of $2$-cells which may be deleted the $Z_2$-Betti number $b_2(G\times G \setminus C)$ is extensive (actually, non-zero suffices for the proof).  Write $X\coloneqq G\times G \setminus C$.  Suppose $X$ is $1$-localized with range $R$ (for all large $N$) and that this property is witnessed by a map $f:X\rightarrow K$, a $1$-complex, i.e. a finite graph.  It is convenient (but not at all essential, as we will see in the proof of Theorem \ref{thm2.5}) to represent a non-zero $2$-cycle in $H_2(X;Z_2)$ by a map $g$ from a (not necessarily orientable) surface $\Sigma$.
$$g:\Sigma\rightarrow X\text{ satisfies:}$$
$$g_*[\Sigma]_2=x\neq 0\in H_2(X;Z_2)$$

Now consider the pre-images $W_p\coloneqq(f\circ g)^{-1}(p)$, $p\in K$, of the composition. We know that for non-empty $W_p$, $\text{diam}(g(W_p))<R$.  At this point to avoid pathologies we will assume (permitting possible subdivision of the cell structure of all spaces $\Sigma$, $X$, and $K$) that $f$ and $g$ are simplicial maps.  If a continuous map $h$ between compact metric spaces has pre-images of diameter $\leq r$ then $\forall\text{ }\epsilon >0$ $\exists\text{ }\delta >0$ such that $\forall$ points $p$, $\text{diam}(h^{-1}(\text{Ball}_{\epsilon}(p)))<r+\delta$.  The simplicial approximation $\bar{f}$ to $f$ can satisfy $\| f-\bar{f}\| _{\text{sup}}<\epsilon$ for any fixed $\epsilon >0$.  Since $\bar{f}^{-1}(p)<f^{-1}(B_\epsilon(p))$, we have $\text{diam}(\bar{f}^{-1}(p))<r+\delta$.  Consequently, restricting to simplicial maps $f$ has no effect, even for the constant, on the property of being ``$R$-localized''.

Let us give the idea of the contradiction we are approaching.  The surface $\Sigma$ is ``foliated'' by ``not-very-large-in-$X$'' sets $W_p$, $\text{diam }g(W_p)<R$.  But
 \[\begin{xy} <5mm,0mm>:
(0,0)*{W_p\xrightarrow{\hspace{0.075in}g\hspace{0.075in}}X\subset G\times G}
,(-2.75,-0.5);(-2.75,-1),**\dir{-}
\ar@{-}_{h} (-2.75,-1);(2,-1)
\ar (2,-1);(2,-0.5)
\end{xy}\]
and the Cartesian geometry $d=\sqrt{d_1^2+d_2^2}$ on $G\times G$ built from the $1$-complex path metric in each factor is locally convex, meaning it has convex neighborhoods---whose radius approaches infinity as $N\rightarrow\infty$---while $R$ is fixed.  This suggests that the map $W_p\rightarrow G\times G$ extends over the mapping cylinder $M$:
\[\xymatrix{
W_p\ar@{^{(}->}[d]_{\operatorname{inc}}\ar[r]^-h&G\times G\\
M\ar@{-->}[ur]^{\bar{h}}
}\]
where $M=\Sigma\times[0,1]/\sigma_1\times 1\equiv \sigma_2\times 1$ whenever $f\circ g(\sigma_1)=f\circ g(\sigma_2)$.  The rough thought is that M is ``foliated'' by cones to various points in $K_0\coloneqq\text{ image }(f\circ g)\subset K$ and that each of these cones separately maps (even canonically to its centroid) into $G\times G$ using the convexity structure.  So far so good, but we would like to state that these cones fit together continuously so that $\bar{h}$ above is a continuous map.  If we had a continuous $\bar{h}$ we would have a contradiction, since the induced diagram (below) on $H_2(\text{ };Z_2)$ shows that $x=0$, whereas by hypothesis $x\neq 0\in H_2(X;Z_2)$.
\[\begin{xy}<5mm,0mm>:
(0,0)*{[\Sigma]\in H_2(\Sigma;Z_2)}="A"
,"A"!UL="A1"
,(8.5,0)*{x\neq 0\in H_2(X;Z_2)}="B"
,"B"!UL="B1"
,"B"!L="B2"
,(20,0)*{y\neq 0\in H_2(G\times G;Z_2)}="C"
,"C"!UL="C1"
,"C"!D="C2"
,"C"!L="C3"
,(0,-3)*{H_2(M;Z_2)}="D"
,"D"!U="D1"
,"D"!D="D2"
,(0,-6)*{H_2(K_0;Z_2)\cong 0}="E"
,"E"!U="E1"
\ar^-{g_*} "A";"B2"+(-0.25,0)
\ar^-{\alpha\coloneqq\text{ inc}_*} "B";"C3"+(-0.25,0)
\ar^{\text{inc}_*} "A";"D1"+(0,0.25)
\ar^{\bar{h}_*} "D";"C2"+(0,-0.25)
\ar^{\|{\large{\wr}}}_{\pi_*} "D2";"E1"+(0,0.25)
\ar@/^18pt/ "A1"+(0.25,0.25);"B1"
\ar@/^20pt/ "B1"+(0.25,0.25);"C1"
\end{xy}\]

Lemma \ref{lem2.3}  and $x\neq 0$ imply $y\coloneqq\alpha(x)\neq 0$, but, in contradiction, the top map factors through $H_2(M;Z_2)\cong 0$.  To see this vanishing, note that any mapping cylinder deformation retracts to its target.  In our case, this means the inclusion $i:K_0\subset M$, induced by $\Sigma\times 1\subset\Sigma\times[0,1]$, has a homotopy inverse $\pi$ induced by the projection $p:\Sigma\times[0,1]\rightarrow\Sigma\times 1$; $\pi\circ i\simeq\text{ id}_{K_0}$, and $i\circ\pi\simeq\text{ id}_M$.  Consequently, $K_0$ and $M$ are homotopy equivalent, $K_0\simeq M$.  By Remark \ref{rem2.4}, since $K_0$ is a $1$-complex, $H_2(K_0;Z_2)\cong 0$, a property inherited by $H_2(M;Z_2)$, since homotopy equivalences induce isomorphisms on homology.

The preceding outline is almost complete.  The flaw is that pre-images of points even under a continuous map can jump discontinuously (consider a not-strictly monotone function from $R$ to $R$ which is constant on exactly one closed interval.  All point pre-images are single points except for that one interval.)  So our proposed ``cone to centroid'' construction might produce discontinuities in $\bar{h}$.  But these are easily repaired.

It was already noted that the pre-images under $f$ of $\epsilon$-balls also satisfy $\text{diam }f^{-1}(B_\epsilon(p))<R$ for sufficiently small $\epsilon>0$.  This allows us to subdivide the target graph $K_0$ so finely that the pre-image of every $1$-simplex (as well as every $0$-simplex) has diameter $<R$.  Now build a slightly different (in topology it would be called the ``block'' version) but homotopy equivalent version $M'$ of the mapping cylinder $M$ and build a map from $M'$ to $G\times G$ by using the local convex structure of $G\times G$ twice.  First cone $(f\circ g)^{-1}$ ($0$-cells) and map these cones linearly to the centroid of $g(f\circ g)^{-1}$ ($0$-cell) $\subset G\times G$.  Then for each $1$-cell $e\subset K_0$ cone: $(f\circ g)^{-1}(e)\cup\text{cone}((f\circ g)^{-1}(\partial_{\verb|_|}e))\cup\text{cone}(f\circ g)^{-1}(\partial_{+}e)$ (Note: the latter two terms of the union and their maps to $G\times G$ were just previously constructed.) and again map this ``secondary'' cone to $G\times G$ linearly to its centroid, $M'$ is easily seen to be homotopy equivalent to $M$.  The easiest way is to construct a deformation retraction $M'\rightarrow K_0$: first compressing the fibers above $0$-cells of $K_0$ and then using Urysohn's Lemma to deform the second cone in a neighborhood of $(f\circ g)^{-1}(\partial e)$ to a mapping cylinder structure as illustrated in Figure 2.1 below.

\begin{figure}[hbpt]
    \[\begin{xy}
        (-50,0)="A"
        ,(-70,0)="A1"
        ,(-30,0)="A2"
        ,(50,0)="B"
        ,(70,0)="B1"
        ,(60,0)="B3"
        ,(30,0)="B2"
        ,(40,0)="B4"
        ,"A"-(0,10)*{\text{original cone of cones}}
        ,"B"+(15,-11)*{\underbrace{}_\text{deform map here}}
        ,"B"+(-15,-11)*{\underbrace{}_\text{deform map here}}
        ,"B"-(0,16)*{\text{final mapping}}
        ,"B"-(0,20)*{\text{cyclindary structure}}
        ,(0,-30)*{\text{FIG. 2.1.  Rotate about }x\text{-axis to make illustration 3D}}
        ,"A"*=<4cm,1.25cm>\hbox{}*\frm{-}*{\bullet}
        ,"B"*=<4cm,1.25cm>\hbox{}*\frm{-}
        ,"A"+(0,6.25);"A"-(0,6.25), **\dir{-}
        ,"A1";"A2", **\dir{-}
        ,"A1"+(0,6.25);"A2"-(0,6.25), **\dir{-}
        ,"A1"-(0,6.25);"A2"+(0,6.25), **\dir{-}
        ,"B"+(0,6.25);"B"-(0,6.25), **\dir{-}
        ,"B3"+(0,6.25);"B4"-(0,6.25), **\dir{-}
        ,"B3"-(0,6.25);"B4"+(0,6.25), **\dir{-}

        \ar "A1"+(0,5);"A1"+(0,1)
        \ar "A1"-(0,5);"A1"-(0,1)
        \ar "A2"+(0,5);"A2"+(0,1)
        \ar "A2"-(0,5);"A2"-(0,1)
        \ar "A1"+(0,6.25);"A"+(-10,3.125)
        \ar "A"+(0,6.25);"A"+(0,3.125)
        \ar "A2"+(0,6.25);"A"+(10,3.125)
        \ar "A2";"A2"-(10,0)
        \ar "A2"-(0,6.25);"A"+(10,-3.125)
        \ar "A"-(0,6.25);"A"-(0,3.125)
        \ar "A1"-(0,6.25);"A"-(10,3.125)
        \ar "A1";"A1"+(10,0)

        \ar "B1"+(0,5);"B1"+(0,1)
        \ar "B1"-(0,5);"B1"-(0,1)
        \ar "B2"+(0,5);"B2"+(0,1)
        \ar "B2"-(0,5);"B2"-(0,1)
        \ar "B1"+(-0.5,6.25);"B1"+(-2.5,1)
        \ar "B1"+(-3.75,6.25);"B1"+(-7,1)
        \ar "B1"+(-7,6.25);"B1"+(-13,1)
        \ar "B1"+(-0.5,-6.25);"B1"+(-2.5,-1)
        \ar "B1"+(-3.75,-6.25);"B1"+(-7,-1)
        \ar "B1"+(-7,-6.25);"B1"+(-13,-1)
        \ar "B2"+(0.5,6.25);"B2"+(2.5,1)
        \ar "B2"+(3.75,6.25);"B2"+(7,1)
        \ar "B2"+(7,6.25);"B2"+(13,1)
        \ar "B2"+(0.5,-6.25);"B2"+(2.5,-1)
        \ar "B2"+(3.75,-6.25);"B2"+(7,-1)
        \ar "B2"+(7,-6.25);"B2"+(13,-1)
        \ar "B3"+(0,6.25);"B"+(5,3.125)
        \ar "B"+(0,6.25);"B"+(0,3.125)
        \ar "B4"+(0,6.25);"B"+(-5,3.125)
        \ar "B4"-(0,6.25);"B"+(-5,-3.125)
        \ar "B"-(0,6.25);"B"-(0,3.125)
        \ar "B3"-(0,6.25);"B"+(5,-3.125)

        \ar@`{(-4,6),(4,-2)} (-10,0);(10,0)
    \end{xy}\]
\end{figure}

This completes the proof of Theorem \ref{thm2.1}.
\end{proof}
\end{theorem}

There are immediate generalizations of Theorem\ref{thm2.1}  in several directions.  One direction is to find other families of spaces, e.g. manifolds with the local geometry of a symmetric space which, for topological reasons, have extensive $Z_2$ homology in some dimension $k$, $\frac{b_k}{volume}\geq\text{constant}$, and to prove that these families are not $(k-1)$-hyperfinite.  But the most straightforward generalization, the one presented next, is to simply consider $k$-fold product of large girth graphs.

\begin{theorem}
\label{thm2.5}
Let $G_N$ be a family of connected degree $=d$ graphs with diverging girth.  The family of $k$ complexes $G_N^k\coloneqq G_N\times\cdots\times G_N$ ($k$-copies) is not ($k-1$)-hyperfinite.

\begin{proof}
The argument is parallel to the proof of Theorem \ref{thm2.1} so we only highlight the differences.

Whereas the second $Z_2$-Betti number $b_2$ was extensive for $G\times G$, an iterated application of the K\"{u}nneth formula shows that the (only) extensive Betti number for $G^k$ is $b_k=(\frac{d-2}{2}N+1)^k$.  A very similar homological argument shows that for $\epsilon >0$ small enough deleting an ``$\epsilon$-fraction'' of $G^k$ results in a subspace $X$ whose $b_k$ is still extensive.

Once $k>2$ an essential cycle $x\in H_k(X;Z_2)$ may no longer be represented by a map of a $k$-manifold, so our surface $\Sigma$ must be generalized to a $k$-complex $\Sigma^k$ with $H_k(\Sigma^k;Z_2)\cong Z_2$ generated by $[\Sigma^k]$ so that the class $0\neq x=g_*[\Sigma^k]$ for some $g:\Sigma^k\rightarrow X$.  The fact that $\Sigma$ was a surface, not a general $2$-complex, actually played no role (other than an aid to visualization) in the proof of Theorem \ref{thm2.1}, so we are not now inconvenienced by needing to take $\Sigma^k$ to be a complex.

Again for contradiction we assume there is a witness to $(k-1)$-hyperfiniteness in the form of a map $f:X\rightarrow K$, to a $(k-1)$-complex.  Again we think of $\Sigma^k$ as ``foliated'' by pre-images $(f\circ g)^{-1}(p)$, $p\in K$, and $\text{diam}(g(f\circ g)^{-1}(p))<R$.  $G^k$ still has a local convex Cartesian geometry so we should attempt to factor the map $\text{inc}\circ g:\Sigma^k\rightarrow G^k$ through $M$, the mapping cylinder.  $\Sigma^k\times[0,1]/(\sigma,1)\equiv(\sigma',1)$ iff $f(\sigma)=f(\sigma')$, as follows:
\[\xymatrix{
\Sigma^k\ar[r]^{g}\ar[d]_{\pi}&X\ar[r]^{\text{inc}}&G^k\\
M\ar@{-->}[urr]^{\bar{h}}\\
}\]
\hspace{2.6in}$\mid${\large{$\wr$}}

\hspace{1.75in}$K_0\subset K$, a $k-1$ complex

All the homological facts are precisely parallel replacing $H_2(\text{ };Z_2)$ by $H_k(\text{ };Z_2)$, specifically since $\text{image }(f)\coloneqq K_0\subset K$ is a ($k-1$)-complex so $H_k(M;Z_2)\cong 0$, again giving a homological contradiction to the existence of $k^\text{th}$ homology in $X$.

In higher dimensions ($k>2$) a bit more needs to be said about the construction of $M'$, the ``block'' model for $M$.  Now instead of two stages of coning there will be $k$ successive stages.  Triangulate $K_0$ finely so that $\text{diam}(f^{-1}\text{ (any cell) })<R$.  Consider first each $f^{-1}$ ($0$-cell) and attach a cone on each of these to $\Sigma^k$.  The map $h\coloneqq \text{inc}\circ g:\Sigma^k\rightarrow G^k$ should be extended over these cones by ``cone to centroid'' using the local convexity structure of $G^k$.  Next consider each $f^{-1}(e)$, $e$ a $1$-cell, and attach a cone to $(f^{-1}(e)\cup\text{ cone }f^{-1}(\partial_{\verb|_|}e)\cup\text{ cone }f^{-1}(\partial_{+}e))$.  Again ``cone to centroid'' extends $h$ over this secondary cone.  Then consider each $f^{-1}(e_2)$, $e_2$ a $2$-cell, and attach a cone to:
$$f^{-1}(e_2)\cup(\bigcup_{e_1\in\partial e_2}\text{ cone}(f^{-1}(e_1)\cup\text{ cone }f^{-1}(\partial_{\verb|_|}e_1)\cup\text{ cone }f^{-1}(\partial_{+}e_1)))$$
\noindent and extend $h$ by ``cone to centroid''.  Continue in this way until each cell $e_k$ of the final stratum is assigned a cone: $\text{cone}(f^{-1}(e_{k})\cup(\text{lower cones}))$, and that cone is mapped via ``cone to centroid'' into $G^k$.  Again an elementary ``bending of cone lines'' shows that this block version $M'$ is also homotopy equivalent to $K_0$ and therefore has vanishing $H_k(M';Z_2)$.  This contradiction completes the proof.
\end{proof}
\end{theorem}

There are numerous reasons detailed below related to efficiency of quantum codes and to qPCP not to be satisfied with product of graph examples but to exploit the full power of Lie theory, Riemannian geometry, and topology to study more exotic contexts in which non-($k-1$)-hyperfinite manifolds arise.  The duality reflected to the toric code fits very clearly with Poincar\'{e} duality in manifolds and makes the search for non-($k-1$)-hyperfinite manifolds, rather than merely complexes, important.  Surprisingly, we find that the mathematical questions that attempts at such constructions raise are just beginning to be addressed in the geometry literature.  In particular, the calculation of $L^2$-cohomology for symmetric spaces has surprising relevance.

Let us take a brief tour of some intriguing classes of examples and record what is known, and what is open for these.

Just as the hyperbolic plane $\mathbb{H}^2$ is in some ways a homogeneous analog of an infinite $d$-valent tree, with $d$ being the analog of the Gaussian curvature, closed hyperbolic surfaces with injectivity radius $r$ approaching infinity (meaning every point $p\in\Sigma$ enjoys an embedded ball neighborhood $B_r(p)\subset\Sigma$ of radius $r$) are analogous to a family of $d$-valent graphs with increasing girth.  (And further, surfaces such as those of arithmetic type, with $\lambda_1$ of the Laplacian bounded away from zero, are analogs of expander graphs.)  By arguments utterly parallel to Theorems \ref{thm2.1} and \ref{thm2.5} we may prove:

\begin{theorem}
Let $\Sigma_A$ be a sequence of compact hyperbolic surfaces with injectivity radius $r$ going to infinity as a function of the area $A$ ($r\approx \frac{2}{3}\log_e A$ is very nearly the optimum).  Then the $k$-fold Cartesian products $\Sigma_\Delta^k$ form a non-($k-1$)-hyperfinite family.

\begin{proof}
Apply the earlier arguments to the $k^\text{th}$ homology.  $b_k(\Sigma_\Delta^k)$ is extensive.  In fact, combining the Gauss-Bonnet theorem and the K\"{u}nneth formula one computes that $\lim_{A\to\infty}b_k(\Sigma^k)/\text{vol}_{2k}(\Sigma^k)=\frac{1}{8\pi}$.  A technical point is that our earlier discussion calibrated Betti number in terms of the number of cells in a complex and now we calibrate in terms of volume.  Fortunately the following theorem of Ref.~\onlinecite{bowditch} allows a direct translation.

\begin{theorem}
Given an integer $d\geq 2$ and a real number $r>0$, there is constant $C(d,r)>0$ so that every hyperbolic $d$-manifold with injectivity radius $>r$ can be triangulated with geodesic $d$-simplicies $\sigma_i$ of bounded geometry in the sense that 
each $\sigma_i$ admits a homeomorphism $h_i:\sigma_i \rightarrow \sigma_0$, $\sigma_0$ the hyperbolic simplex with all sides of length $1$ so that
$$
\frac{1}{C(d,r)} d(x,y) \leq d(h_i(x),h_i(y) \leq C(d,r) d(x,y)
$$
for all $x,y\in \sigma_i$.
\begin{proof}

See Ref.~\onlinecite{bowditch}.
\end{proof}
\end{theorem}

Given this theorem, the modification from the proof of Theorems \ref{thm2.1} and \ref{thm2.5} is routine.
\end{proof}
\end{theorem}

According to the Chern-Weil theory, the Euler characteristic $\chi$ of a Riemannian manifold may be computed by integrating the Pfaffian of the Levi-Civita curvature over the manifold.  For closed odd-dimensional manifolds the result is zero and for hyperbolic $2n$-manifolds $M^{2n}$ a local expression shows that there are dimension-dependent constants $c_n$ so that the Euler characteristic satisfies:
$$\chi(M)=c_n\text{vol}(M).$$
$c_1=-\frac{1}{2\pi}$ and in general the $c_\text{odd}$ are negative and the $c_\text{even}$ positive\cite{EulerCharacteristic}.

Since there is also a combinatorial formula for $\chi(M)$ in terms of the real (equivalently rational, $Q$) Betti numbers:
$$\chi(M)=\Sigma^{2n}_{i=0}(-1)^i b_i^Q.$$
We learn that certain sums must be extensive.  For example, we have this table for closed hyperbolic $2n$-manifolds:

\begin{table}[hbtp]
    \begin{center}
        \begin{tabular}{|c|c|}
            \hline
            dimension $=2n$ &   extensive quantities    \\
            \hline\hline
            $2$             &   $b_1^Q$                 \\
            \hline
            $4$             &   $b_2^Q$                 \\
            \hline
            $6$             &   $b_1^Q+b_3^Q+b_5^Q$, $b_1^Q+b_3^Q$  \\
            \hline
            $8$             &   $b_2^Q+b_4^Q+b_6^Q$, $b_2^Q+b_4^Q$  \\
            \hline
        \end{tabular}
    \end{center}
\end{table}

(Poincar\'{e} duality implies $b_k^Q=b^Q_{2n-k}$, accounting for the multiple extensive quantities listed.)

In all our application either $O(1)$-depth quantum circuits or length scales protecting quantum information arise so we are really only interested in toric codes on sequences of manifolds whose injectivity radius diverges.  In this case there is more refined information\cite{gromov,bergeron} coming from the behavior of $L^2$-cohomology under the processes of taking geometric limits of symmetric spaces.

\begin{theorem}
If $\{M^d\}$ is a sequence of hyperbolic $d$-manifolds, $d$ fixed, with injectivity radius diverging, then all densities $b_i^Q/\text{vol}\to 0$ except $b_n^Q/\text{vol}$, which approaches positive universal values for $d=2n$.
\end{theorem}

In building quantum codes $\zc$ Betti numbers are more relevant than rational ones; the toric code is based, for example, on $\zz$-homology.  Fortunately the universal coefficient exact sequence for arbitrary X:
$$H_i(X;Z)\otimes \zc \longrightarrow H_i(X;\zc)\longrightarrow\text{ Tor}(H_{i-1}(X;Z),\zc)\longrightarrow 0$$
\hspace{4.25in}$\|${\large{$\wr$}}

\hspace{3.25in}$\text{torsion}(H_{i-1}X;Z)\otimes \zc$

together with the identity
$$\text{rank}(\text{Free}(H_i(X;Z)))=b_i^Q(X)$$
\noindent shows that the $\text{mod } \coeff$ Betti numbers satisfy:
$$b_i^{\zc}(X)=\text{rank}(H_i(X;\zc))\geq\text{rank}(\text{Free}(H_i(X;Z)))=b_i^Q(X).$$
So $b_i^Q$ serves as a lower bound for all $b_i^{\zc}$.

Again a straightforward generalization of the proofs of Theorems \ref{thm2.1} and \ref{thm2.5} establishes

\begin{theorem}
Any sequence $\{M\}$ of hyperbolic $2n$-manifolds with injectivity radius diverging is non-($n-1$)-hyperfinite.
\end{theorem}

Note that all complex matrix groups are residually finite, and by Selberg's lemma each admits torsion-free subgroups of finite index.  These two facts permit the construction of a sequence of hyperbolic manifolds with diverging injectivity radius within the covering space tower of any fixed hyperbolic manifold.

Although we now have excellent control of rational Betti numbers we really only have lower bounds on $\zc$ Betti numbers.  We have an:

\begin{openquestion}
\label{oq2.10}
For $\{M\}$ a sequence of hyperbolic $d$-manifolds, $d$ fixed $=3,4,5,6,\ldots$, with injectivity radius diverging, is it possible that a Betti number $b_i^{\zc}$, for $i\neq\frac{d}{2}$ could have nonzero density in the limit?
\end{openquestion}

\noindent{\underline{Discussion:}}
This question is number theoretic in nature.  Investigation of analytic pro-$p$ groups suggests some scaling like $b_1^{\zc}\approx\text{vol}^{\frac{1}{2}}$ might be achieved for certain sequences at arithmetic $3$-manifolds (with diverging injection radius).

Curiously, convincing heuristics and copious numerical evidence\cite{heurnum} support the conjecture that for certain sequences of arithmetic hyperbolic $3$-manifolds (with diverging injection radius),
$$\log_e(\text{order torsion}(H_1(M;Z))/\text{vol}\to\frac{1}{6\pi},$$
$\frac{1}{6\pi}$ being the $L^2$-analytic torsion of $\mathbb{H}^3$.

Unfortunately, the numerics show that many primes and high prime powers factor this order so this observation does not imply extensive $b_1^{\zc}$ for any $\coeff$.
See the discussion for the relevance of this open question.

\section{Quantum As the Square of Classical}
\label{QSqCl}
Before launching into the definitions and constructions, we give some additional motivation for this problem.  As noted, for the qPCP conjecture to be true, we need a family of Hamiltonians with NLTS.  Sometimes, the property that a state is not trivial is referred to as a state being ``topologically ordered", although often the term ``topological order" is used in a slightly different sense.  Let us return to the classical setting and consider the property of long-range order: a long-range correlation between two degrees of freedom in the ground state.  We begin with one-dimensional systems with long-range order in the ground state, but not at thermal equilibrium in non-zero temperature.  Then, we turn to two-dimensional systems, with long-range order at sufficiently small non-zero temperature, but which have low energy states without long-range order, and finally we discuss these systems on expander graphs, showing that every low energy density state has long-range order in a certain sense (this property may be known elsewhere but we do not know a reference to it in the literature).
We then give a parallel discussion in the quantum case; the graphs that we consider in the quantum case are precisely the square of the graphs we considered in the classical case.  We note that the two-dimensional toric code has a nontrivial ground state, but has trivial states at nonzero temperature.  We then turn to a four dimensional-toric code which has topological order for sufficiently small non-zero temperatures but which has low energy states which are trivial.  This then motivates a consideration of a toric code on the square of an expander graph, which we consider in the next section, section \ref{toward}.

Consider the properties of the classical ferromagnetic Ising model on various graphs.  To define this Ising model, for each graph $G$, we have a spin-$1/2$ degree of freedom on every vertex.  The Hamiltonian is
\be
\label{HIsingFerro}
H_{Ferro}=-\sum_{<i,j>} S^z_i S^z_j,
\ee
where $\sum_{<i,j>}$ denotes a sum over all pairs of vertices $i,j$ which are neighbors on the graph.
If the graph is a one-dimensional line, then the Ising model has long-range order in the ground state: while there is a two-dimensional space of ground states, either all spins up or
all spins down, the expectation value $\langle \psi, S^z_i S^z_j \psi \rangle$ is equal to plus $1$ in all ground states $\psi$, for all sites $i,j$, regardless of whether or not $i,j$ are neighbors.
However, the long-range order in this case is not stable to non-zero temperature.  Let $\rho(T)=\exp(-H/T)/{\rm tr}(\exp(-H/T))$, so that
$\rho$ is the thermal density matrix at any temperature $T>0$.  The expectation value ${\rm tr}(\rho(T) S^z_i S^z_j)$ decays exponential in the distance between $i$ and $j$ for any $T>0$.  However, if we move to a two-dimensional Ising model, the long-range order does survive at small enough non-zero temperature.  Consider a families of graphs $G_L$, where $G_L$ is a square-lattice of size $L$-by-$L$, and let $\rho_L(T)$ be the corresponding thermal density matrix for graph $G_L$.  Then, there is a $T_c>0$ such that for all $T<T_c$ there is a quantity $m(T)>0$ such that ${\rm tr}(\rho_L(T) S^z_i S^z_j)\geq m(T)$ for all $L$ and all vertices $i,j$ in $G_L$.

However, while the long-range order survives at non-zero temperature in the two-dimensional (and higher) model, there are states $\psi_{low}$ of arbitrary low energy density which are not long-range ordered.  We define the energy density of a state as follows:
\begin{definition}
Consider a Hamiltonian $H$ written as a sum of $N_{terms}$ different terms, $H=\sum_{i=1}^{N_{terms}} H_i$, where $\Vert H_i \Vert \leq 1$ for all $i$.
Given any density matrix $\rho$, we define the {\bf energy density} of $\rho$ to be
\be
\frac{{\rm tr}(\rho H)-E_0}{N_{terms}},
\ee
where $E_0$ is the lowest eigenvalue of $H$.
\end{definition}
Note that the energy density may depend upon the decomposition used to write $H$.

A simple way to construct $\psi_{low}$ is as follows.  Take the $L$-by-$L$ square lattice and divide it into small squares of size $l$-by-$l$, for $l<<L$ (if $L/l$ is not an integer, we can use a mix of squares and rectangles of size $l$ and $l+1$).  Label the small squares by an index $a$ and consider the state $\rho=\prod_{a} \rho_a$, where $\rho_a$ is a mixed state: it is the equal probability mixture of the state with all spins up in square $a$ and the state with all spins down, so $\rho_a=(1/2)(|\uparrow\uparrow\uparrow...\rangle\langle \uparrow \uparrow\uparrow ...|+|\downarrow\downarrow\downarrow...\rangle\langle\downarrow\downarrow\downarrow...|)$.
Then, this state $\rho$ has an energy density of order $l/L$ which can be made arbitrarily small for large $L$, but the correlation function ${\rm tr}(\rho S^z_i S^z_j)$ is equal to $0$ is sites $i,j$ are in different small squares.

We could define long-range order in this case in a slightly different way.  Consider the quantity
\be
M^2=\frac{1}{N} {\rm tr}\Bigl( \rho (\sum_i S^z_i)^2 \Bigr)^2,
\ee
where $N=L^2$ is the total number of spins.  Using the checkerboard state defined above, we have $M^2=0$ if $L/l$ is an even integer.  For the mixed state $\prod_a \rho_a$, we have $M^2 \sim l^2/L^2<<1$.  Note that this quantity $M^2$ is equal to the average, over all choices of $i,j$, of the correlation function ${\rm tr}(\rho S^z_i S^z_j)$.  On expanders (below), $M$ behaves quite differently.

Note that nothing in the above discussion of the checkerboard state or state $\prod_a \rho_a$ relied crucially on having a two-dimensional lattice.  We could have made a similar construction using small hypercubes of linear size $l$ for any finite dimensional lattice.

So, finally, we consider an Ising model on an expander graph.
We define the expansion coefficient by
\begin{definition}
Suppose the graph has $N$ vertices and has the expansion property that for every set $S$ of vertices with $|S|\leq N/2$, the number of edges in the graph connecting $S$ to its complement has cardinality at least $c |S|$ for some constant $c>0$.  Then the graph has {\bf expansion coefficient} $c$.
\end{definition}
Suppose $c>0$.
Then, we claim for any state of sufficiently low energy, the quantity $M^2$ is of order unity, in contrast the the case on the square lattice.  Consider a configuration of spins with $N_\uparrow$ spins up (that is, with $S^z_i=+1$) and $N_\downarrow$ spins down with $N_\uparrow+N_\downarrow=N$.  Assume that $N_\uparrow \geq N_\downarrow$.  Then the set of vertices with down spins has cardinality at most $|S|/2$ and hence the number of edges connecting the down spins to the up spins is at least $c N_\downarrow$.  In general, we find that the number of edges connecting down spins to up spins is at least ${\rm min}(N_\uparrow,N_\downarrow)$.  Let the energy of the ground states of the system (either the state with all spins up or the state with all spins down) be $E_0$.  Note that $E_0$ is equal to $-1$ times the number of edges.  Every edge connecting an up spin to a down spin raises the energy by $2$ compared to the ground state energy so
the energy $E$ of a configurations with $N_\uparrow$ up spins and $N_\downarrow$ down spins obeys
\be
E \geq E_0+2 c{\rm min}(N_\uparrow,N_\downarrow).
\ee
Note that
\be
M^2=\frac{(N_\uparrow-N_\downarrow)^2}{N^2}.
\ee
Assuming without loss of generality that $N_\downarrow \leq N_\uparrow$, we find that
\begin{eqnarray}
M^2&=&(N-2N_\downarrow)^2/N^2
\\ \nonumber
&\geq & (N^2-4 N N_\downarrow)/N^2
\\ \nonumber
& = &1-4N_\downarrow/N
\\ \nonumber
&\geq & 1-\frac{2(E-E_0)}{cN}.
\end{eqnarray}
For a mixed state $\rho$, we can average both sides of the above equation and find that
\be
\label{M2en}
M^2 \geq 1-\frac{2(E-E_0)}{cN},
\ee
where now $E={\rm tr}(\rho H_{Ferro})$.
Note that the number of terms in $H_{Ferro}$ is equal to $(N/2)$ times the degree $d$ of the graph.  So,
$(E-E_0)/N$ is equal to the energy density times $2/d$, and so indeed states with small enough energy density have $M^2$ of order unity.

We now turn to the case of topological order, considering the toric code on various lattices.  First consider the toric code on a two dimensional lattice.
Depending on boundary conditions chosen (periodic, i.e. torus boundary conditions, open boundaries, etc...) there may be one or more zero energy ground states.  However, these states are nontrivial under the circuit definition, in the sense that none of them is $(R,\epsilon)$-trivial for $R/L$ sufficiently small and for $\epsilon$ sufficiently small compared to unity.  See for example Ref.~\onlinecite{bhv} for a torus geometry or Ref.~\onlinecite{leshouches} for other geometries.  However, at any fixed nonzero temperature the states are trivial.  More precisely, for any such temperature, for any $L$, the state is $(R,1/2)$ trivial for an $R$ which grows only logarithmically with $L$ (the prefactor in this logarithmic dependence depends upon temperature).  See Ref.~\onlinecite{mixedtriv}, for example.

Turning to the case of the four-dimensional lattice, there are various possible toric codes that can be defined.  In general, in $d$ dimensions, we can refer to a $(q,r)$ code, with $q+r=d$, where $q-1$ is the dimensionality of electric defects and $r-1$ is the dimensionality of magnetic defects.  The electric defects bound $q$-dimensional surfaces, and the magnetic defects bound $r$-dimensional surfaces.  The toric code discussed above in two dimensions is an example of a $(1,1)$ code.  In the next section, we describe a generalization of this toric code to arbitrary two-complexes.  However, for now we consider a $(2,2)$ code on the four-dimensional lattice.  In this case, for sufficiently small nonzero temperature, the state is not $(R,\epsilon)$ trivial for $R/L$ sufficiently small and $\epsilon$ sufficiently small compared to unity\cite{4dtoric,mixedtriv}.

So, we have seen three classical systems, one of which has long-range order only at zero temperature, the next has long-range order at sufficiently small temperature, and the third has order for all sufficiently low energy states.  We have quantum systems that are in a sense analogous to the first two of these.  The question naturally arises then: is there a quantum system that has NLTS?  This is the question we address.

\section{Toric Codes on Complexes}
\label{tconcomplex}
For any $2$-complex, we can define a Hamiltonian that generalizes the toric code Hamiltonian as follows.  Associated with every $1$-cell of the complex is a spin-$1/2$ degree of freedom.  The Hamiltonian is a sum of terms
\be
\label{Htc}
H_{tc}=A+B,
\ee
where $A,B$ are defined by
\be
\label{Adef}
A=-\sum_{s} A_s,
\ee
\be
\label{Bdef}
B=-\sum_p B_p.
\ee
The sum in Eq.~(\ref{Adef}) is a sum over all $0$-cells, where $s$ labels the $0$-cell, and $A_s$ is defined to by
\be
A_s=\prod_{s \in \partial e}S^x_e,
\ee
with the product being a product over all $1$-cells $e$ that are attached to $s$ (so that $s$ is in the boundary $de$ of $e$) and $S^x_e$ is the spin operator $S^x$ acting on the degree of freedom on $e$.
The sum in Eq.~(\ref{Bdef}) is a sum over all $2$-cells, where $p$ labels the $2$-cell, and $B_p$ is defined to by
\be
B_p=\prod_{e\in \partial p}S^z_p,
\ee
with the product being a product over all $1$-cells $e$ that are attached to $p$ and $S^z_e$ is the spin operator $S^z$ acting on the degree of freedom on $e$.
We have $[A_s,B_p]=0$ for all $s,p$, for all $2$-complexes.

We refer to the terms $A_s$ as vertex terms and the terms $B_p$ as plaquette terms.  If there is a state $\psi$ and a  $0$-cell $s$ for which $A_s \psi=-\psi$, we say that $\psi$ has a vertex defect.  The number of vertex defects of $\psi$ is given by $\sum_s \langle \psi, (1-A_s) \psi \rangle/2$.  We use similar terminology for plaquette defects.  The ground state has expectation value of $A$ equal to $-N_v$, where $N_v$ is the number of $0$-cells and it has expectation value of $B$ equal to $-N_p$, where $N_p$ is the number of $2$-cells.

The construction above is given for an arbitrary complex.  If the complex is a triangulation of a manifold, then the above code is a $(1,d-1)$ code.

\section{Toric Code on $G\times G$: Toward the Quantum PCP Conjecture}
\label{toward}

We now consider this Hamiltonian (\ref{Htc}) on the $2$-complex obtained from $G\times G$ defined previously.  The graphs $G$ will all have bounded degree and diverging girth.
We will prove a ``one-sided" result on the absence of low energy trivial states with no vertex defects, theorem \ref{noletriv}.
Later, in section \ref{cobound} we discuss a possible alternative proof idea of our result; this alternative proof idea relies on an unproven conjecture but may be interesting as, if this conjecture is proven, this method may be useful for a variety of other Hamiltonians.

\begin{definition}
Give a set $S$ of Hermitian operators which square to the identity, we say that a state $\psi$ is {\bf stabilized} by $S$ if $P\psi=\psi$ for all operators $P$ in $S$.
\end{definition}
Note that in this definition that $S$ may be an arbitrary set of such operators.  There is a formalism called the stabilizer formalism based on operators which are products of Pauli operators, but we do not make this restriction here.  We also do not require that the operators commute with each other.

Suppose $\psi$ is a trivial state.  Then, $\psi=U_{circuit} \xi$ for some product state $\xi$ and for some unitary $U_{circuit}$ defined by a local quantum circuit.
The product state $\xi$ is stabilized by a set of operators $O_i$, one such projector for every degree of freedom $i$.  Each such operator $O_i$ is a tensor product of an operator on degree of freedom $i$ with the identity operator elsewhere; the operator on site $i$ has a unique eigenvalue equal to $+1$, so that specifying that $\psi$ is stabilized by the set of operators $O_i$ uniquely specifies $\psi$ up to a scalar.  In the case of the toric code, since the degrees of freedom are on $1$-cells, the index $i$ ranges over all $1$-cells.
Therefore, $\psi$ is stabilized by the set of operators $U_{circuit} O_i U_{circuit}^\dagger$.  Define $\tilde O_i=U_{circuit} O_i U_{circuit}^\dagger$.
Note that if the quantum circuit has range $R$, then each operator $\tilde O_i$ has support on the degrees of freedom within distance $R$ of $i$.
If $\psi$ minimizes $A$, then $A_s \psi=\psi$ for all $s$, and so $\psi$ is stabilized also by the set of operators $A_s$.

We now use this idea of stabilizers to define a new state $\phi$ in lemma \ref{phidefn}.  Very roughly speaking, $\phi$ will have a strange property: in local regions it will ``look like" $\psi$ does (that is, have the same reduced density matrix), up to possible conjugation by a certain unitary operator that anti-commutes with certain plaquette operators $B_p$.  Thus, the unitary operator can be viewed as ``inserting defects" on the plaquettes that it anti-commutes with.  However, there will be a seemingly contradictory property.  To explain this property, note that we are used to the requirement that defects be inserted in pairs in the two-dimensional toric code on a torus, and every state must have an {\it even} number of plaquette defects (this is a consequence of the second homology of the torus), and in this $G \times G$ there is a similar requirement: due to the branching structure of $G \times G$, things are more complicated but still not all configurations of defects are possible.  However, the conjugation by unitary that we use to define $\phi$ will appear not to respect this property: it will seem to add defects in certain places but not add corresponding defects in other places.  The resolution of this apparent paradox will turn out to be that the original state $\psi$ had defects, and in fact we will use this construction of $\phi$ to prove that $\psi$ has defects, and indeed to prove a lower bound on the energy density of $\psi$ for the toric code on $G \times G$.

Before defining the state $\phi$, let us define the needed unitary operators.
The definition of these operators is fairly complicated so before giving the definition, we will explain the goal of the definition by analogy with the two dimensional toric code.  In the two dimensional toric code, the product of $\sigma^x$ along a line of bonds on the dual lattice creates a pair of anyons at the ends of the line (i.e., thinking of the line as a $1$-chain, it creates anyons in the coboundary).  Of course, the product of $\sigma^x$ around a closed loop on the dual lattice commutes with the Hamiltonian but here we are considering a line with endpoints.
There are various possible choices of lines homologous to the given on the dual lattice line, and for each such line we can define an operator which is the product of $\sigma^x$ along that line.  Acting on the ground state, all of these operators product the same excited state.
In the toric code on $G \times G$, we can create a similar product of $\sigma^x$ which will create anyon excitations.  However, due to the branching nature of $G \times G$, the construction of this operator is more complicated to define.  Also, while the operator for the toric code in two dimensions created just a pair of defects at the ends of a line, the operator we define here will create many defects rather than just two.  We will construct the operator so that it creates one defect on a plaquette $p$ and all of the other defects are located distance $g/2$ from $p$.  We will construct two different operators that construct the same set of defects, calling these operators $C^x(p)$ and $D^x(p)$; these two different operators are analogous to two homologous choices of lines in the two-dimensional code.
\begin{definition}
Let $p$ be any $2$-cell in $G\times G$. Define the operator
$D^x(p)$ as follows (this construction involves an arbitrary choice of a $1$-cell below; this choice can be made in any way, but every time we refer to the operator $D^x(p)$ for given $p$, we take the same choice).
 A $1$-cell in $G \times G$ is labeled by an edge in one graph and a vertex in another graph.  Pick a $1$-cell attached to $p$ which is
 labeled by an edge $a$ in the first graph and a vertex $i$ in the second graph.  So, we refer to this $1$-cell as $(a,i)$.
Pick another $1$-cell attached to $p$ which is  labeled by an edge $b$ in the second graph and a vertex $j$ in the first graph.  So, we refer to this $1$-cell as $(j,b)$.  Note that edge $a$ is attached to vertex $j$ in $G$ and edge $b$ is attached to vertex $i$ in $G$.

If the graph $G$ has degree $d$, then $i$ has $d-1$ edges attached to it other than $b$.  Let $R$ be the set of all vertices in $G$ which are within distance $g/2$ of $i$, where $g$ is the girth of $G$, such that the shortest path from $i$ to the given vertex does not use edge $b$.  There are $(d-1)+(d-1)^2+...+(d-1)^{g/2}$ such vertices in $R$.

Define the operator $D^x(p)$ to be the product of $\sigma^x_e$ over all $1$-cells $e$ of the form $(a,k)$ for $k$ in $R$.

We define another operator $C^x(p)$ as follows.  Define $T$ to be the set of all vertices in $G$ which are within distance $g/2$ of $j$,
such that the shortest path from $j$ to the given vertex does not use edge $a$.  Let $\partial R$ be the boundary of $R$: the set of edges in $G$ which join vertices in $R$ to vertices not in $R$; let $\partial T$ be similarly defined as the boundary of $T$.
Define $C^x(p)$ to be the product of $\sigma^x_e$ over all $1$-cells $e$ of the form $(k,b)$ for $k$ in $T$ and over all $1$-cells $e$ of the form $(c,k)$ for $k$ in $R$ and $c$ in $\partial T$ and over all $1$-cells $e$ of the form $(k,c)$ for $k$ in $T$ and $c$ in $\partial R$.
\end{definition}

We note that the product $C^x(p) D^x(p)$ is equal to a product of operators $A_s$ over an appropriately chosen set of sites $s$.  That is, the two $1$-chains corresponding to these two different operators are homologous as claimed above.
Further, $C^x(p)$ anti-commutes with $B_p$, as well as with certain other plaquette operators a distance $g/2$ from $p$ (the plaquette operators it anti-commutes with are those in the coboundary of the $1$-chain corresponding to $C^x(p)$).  It commutes with all other $B_p$.
The operator $D^x(p)$ has the same commutation and anti-commutation relations with $B_p$.

We now give lemma \ref{phidefn}.   Again before giving it we explain by analogy with the two dimensional toric code.  In this analogy, let us consider the two dimensional toric code on a sphere.  Suppose we had a trivial state $\psi$ that had no vertex defects on the sphere.  This state will be stabilized by some operators $\tilde O_i$ as above.  Let $C$ be an operator that anti-commutes with the plaquette operators at the north and south pole of the sphere; i.e., it will be a product of $\sigma^x_i$ along some line $l_C$ on the dual lattice from the north to the south pole.  Let $D$ be a product of $\sigma^x_i$ along some other line $l_D$ on the dual lattice from the north to the south pole, choosing the lines so that $l_C$ and $l_D$ are far away from each other except near the north and south poles (for example, choose both lines to be along longitude lines of the sphere at opposite longitudes).  Then, $C \psi$ is also a trivial state, and is stabilized by $C \tilde O_i C$.  However, since $\psi$ has no vertex defects, $C \psi = D \psi$ so $C\psi$ is also stabilized by $D \tilde O_i D$.  Thus, we claim that $C \psi$ is stabilized by $\tilde O_i$ for $\tilde O_i$ supported far from either the north or south pole (i.e., near the equator).  To see this, consider any such $\tilde O_i$.  If the support of $\tilde O_i$ does not intersect $l_C$, then $C \tilde O_i C = \tilde O_i$ so the claim follows, while if the support of $\tilde O_i$ does intersect $l_C$ then the support of $\tilde O_i$ does not intersect $l_D$ so $D \tilde O_i D = \tilde O_i$ so again the claim follows.  
So, $C\psi$ is stabilized by the same stabilizers $\tilde O_i$ as before for $\tilde O_i$ whose support is far from the north or south pole, while the stabilizers are changed near the north and south pole.  
This allows us to define a new state $\phi$ whose stabilizers agree with $\tilde O_i$ for $\tilde O_i$ far from the south pole, changing the stabilizers only near the south pole but {\it not} near the north pole; we do this by replacing the stabilizers of $C \psi$ near the north pole by the stabilizers of $\psi$ near the north pole.  
Some care is required to show that this replacement can be done consistently, as described below.  The result is a state $\phi$ whose reduced density matrix is the same as that of $\psi$ away from the south pole, while near the south pole the reduced density matrix is the same as that of $C\psi$; that is, we have ``inserted a defect near the south pole", without inserting one elsewhere.  
This provides one way to prove that the state $\psi$ must have plaquette defects and cannot be a ground state of the two-dimensional toric code (the arguments in Refs.~\onlinecite{bhv,leshouches} are much simpler): if $\psi$ indeed had no plaquette defects, then $\phi$ would have an odd number of plaquette defects which is impossible, giving a contradiction.  The resolution of the apparent paradox is that in fact $\psi$ must have been a superposition of states with different numbers of plaquette defects.
In lemma \ref{phidefn} we will want to insert defects in many places, not just one, so this will also require more care.

\begin{lemma}
\label{phidefn}
Let $\psi$ be a trivial state stabilized by the set of operators $\tilde O_i$ as defined above.  Let $\psi$ also be stabilized by the set of operators $A_s$.
Let ${\cal P}$ be any set of plaquettes.
Assume that the range $R$ of the quantum circuit is less than $g/32$.
Then, there exists a state $\phi$, with $|\phi|=1$, such that
$\phi$ is stabilized by the set of operators $A_s$ and also is stabilized by the set of operators
\be
\label{stabset}
\Bigl(\prod_{p\in {\cal P}, {\rm dist}(p,i) \leq g/4} D^x(p) \Bigr)
\tilde O_i
\Bigl(\prod_{p\in {\cal P}, {\rm dist}(p,i) \leq g/4} D^x(p) \Bigr),
\ee
where the notation on the product means the product over all $p$ in ${\cal P}$ such that ${\rm dist}(p,i)\leq g/4$.  Note that the operator $\Bigl(\prod_{p\in {\cal P}, {\rm dist}(p,i) \leq g/4} D^x(p) \Bigr)$ is a Hermitian unitary.

Further, consider any set $X$ of $1$-cells in $G\times G$, with diameter of $X$ at most $g/8$.  Let ${\cal P}_X$ be the set of plaquettes in ${\cal P}$ such that all $1$-cells attached to the plaquette are in $X$.  Let $\phi_X$ be the reduced density matrix of the state $\phi$ on $X$ and let $\psi_X$ be the reduced density matrix of the state $\psi$ on $X$.
Then,
\be
\label{local}
\phi_X=\Bigl(\prod_{p\in {\cal P}_X} D^x(p) \Bigr) \psi_X \Bigl(\prod_{p\in {\cal P}_X} D^x(p) \Bigr).
\ee
\begin{proof}
We show Eq.~(\ref{local}), showing the equation under the assumption that $\phi$ exists.  Then we prove the existence of $\phi$.

First, we note an important property.  There is an arbitrariness in how we chose the stabilizers.  For each $i$, define an arbitrary set of $0$-cells called $S_i$.
Then, if a state is stabilized by
the set of operators $A_s$ and also is stabilized by the set of operators
\be
\label{stabset2}
\Bigl( \prod_{s \in S_i} A_s \Bigr)
\Bigl(\prod_{p\in {\cal P}, {\rm dist}(p,i) \leq g/4} D^x(p) \Bigr)
\tilde O_i
\Bigl(\prod_{p\in {\cal P}, {\rm dist}(p,i) \leq g/4} D^x(p) \Bigr)
\Bigl( \prod_{s \in S_i} A_s \Bigr),
\ee
then it is also stabilized by the operators in Eq.~(\ref{stabset}) and conversely if it is stabilized by the set of operators $A_s$ and also is stabilized by the set of operators in Eq.~(\ref{stabset}) then it is also stabilized by the set of operators in Eq.~(\ref{stabset2}).
This can be regarded as a ``gauge freedom".
Since, as we have noted before, the product $C^x(p) D^x(p)$ is a product of $A_s$ over an appropriate set of $s$, this means that we can replace operators $D^x(p)$ in Eq.~(\ref{stabset}) by operators $C^x(p)$ as desired, so long as we replace all occurences of $D^x(p)$ for a given $i,p$.
That is, for each $i$, we can define a set ${\cal D}_i$ with ${\cal D}_i \subset {\cal P}$ and then the state can be taken to be stabilized by the operators
\be
\label{stabset3}
\Bigl(\prod_{p\in {\cal D}_i, {\rm dist}(p,i) \leq g/4} D^x(p) \Bigr)
\Bigl(\prod_{p\in {\cal P}\setminus {\cal D}_i, {\rm dist}(p,i) \leq g/4} C^x(p) \Bigr)
\tilde O_i
\Bigl(\prod_{p\in {\cal P}\setminus {\cal D}_i, {\rm dist}(p,i) \leq g/4} C^x(p) \Bigr)
\Bigl(\prod_{p\in {\cal D}_i, {\rm dist}(p,i) \leq g/4} D^x(p) \Bigr).
\ee

Consider a given $i$.
Note that if ${\rm dist}(p,i) \leq g/4$ and the operator $\tilde O_i$ is supported on a set of diameter at most $R$, then one of the following cases holds.
Either, $[D^x(p),\tilde O_i]=0$ or $i$ is within distance $R$ of the support of $D^x(p)$.  However, this latter case means that either $i$ is not within distance $R$ of the support of $C^x(p)$ or $i$ is within distance $R$ of $p$.  Thus, there are three possible cases if ${\rm dist}(p,i) \leq g/4$: $\tilde O_i$ commutes with $D^x(p)$, or $\tilde O_i$ commutes with $C^x(p)$, or $i$ is within distance $R$ of $p$.  If $\tilde O_i$ commutes with $D^x(p)$ then we can remove all occurences of $D^x(p)$ from Eq.~(\ref{stabset3}).  If $\tilde O_i$ commutes with $C^x(p)$, then we can conjugate Eq.~(\ref{stabset3}) by $C^x(p)$, and then conjugate by a product of operators $A_s$ to turn that $C^x(p)$ into a $D^x(p)$.
Thus, for any $i$, we can define any set of plaquettes ${\cal P}_i$ such that every plaquette $p$ in ${\cal P}_i$ has ${\rm dist}(p,i)\leq g/4$ and such that every plaquette $p$ in ${\cal P}$ with ${\rm dist}(p,i) \leq R$ is in ${\cal P}_i$, and then the state is stabilized by the set of $A_s$ and by the set of operators
\be
\label{stabsetfinal}
\Bigl(\prod_{p\in {\cal P}_i} D^x(p) \Bigr)
\tilde O_i
\Bigl(\prod_{p\in {\cal P}_i} D^x(p) \Bigr).
\ee
This is the final form of the stabilizers that we will take.
One way to describe these various choices of stabilizers is to note that the stabilizers generate a group and every element of the group stabilizes $\psi$: these different choices of the stabilizers correspond to conjugating certain generators of the group by other generators of the group, giving a different presentation of the same group.
However, now Eq.~(\ref{local}) follows almost immediately.  For each $i$ such that ${\rm dist}(i,X) \leq R$, take ${\cal P}_i$ to include all plaquettes $p \in {\cal P}$ within distance $2R$ of $X$.  Thus, ${\cal P}_i$ has the needed property of including all plaquettes within distance $R$ of $i$.
Also, by a triangle inequality, since $i$ is within distance $R$ of $X$, every plaquette in ${\cal P}_i$ is within distance $3R+g/8$ of $i$, and for the given choice of $R$ this is less than $g/4$.

We now prove that such a state $\phi$ exists.  The proof is inductive.  The induction is on the cardinality of ${\cal P}$.  It is clear that the claim holds when ${\cal P}$ is the empty set, since then we can take $\phi=\psi$.  So, we must prove that if the claim holds for a given set ${\cal P}$, then it also holds for the set of plaquettes ${\cal P}' \equiv {\cal P} \cup \{q\}$, for any plaquette $q$.  We now show this.
Assume that such a $\phi$ exists for a given set ${\cal P}$.
Thus, there is a state stabilized by the set of operators $A_s$ and by the set of operators
\be
\label{stabset5}
\Bigl(\prod_{p\in {\cal P}, {\rm dist}(p,i) \leq g/4} D^x(p) \Bigr)
\tilde O_i
\Bigl(\prod_{p\in {\cal P}, {\rm dist}(p,i) \leq g/4} D^x(p) \Bigr).
\ee
Hence there is a state stabilized by the operators $A_s$ and by the set of operators in Eq.~(\ref{stabset5}) for ${\rm dist}(q,i)>R$.
(This last statement is a triviality: since it is stabilized by the operators of Eq.~(\ref{stabset5}), it is stabilized by a subset of those operators).

Note, then, that there is a state stabilized by the set of operators
$A_s$ and by the set of operators
\be
\label{stabset6}
\Bigl(\prod_{p\in {\cal P}', {\rm dist}(p,i) \leq g/4} D^x(p) \Bigr)
\tilde O_i
\Bigl(\prod_{p\in {\cal P}', {\rm dist}(p,i) \leq g/4} D^x(p) \Bigr)
\ee
for ${\rm dist}(q,i)>R$.
Here we have used the fact that for ${\rm dist}(q,i)>R$ and ${\rm dist}(q,i)\leq g/4$, we can choose to conjugate a given stabilizer by the operator $D^x(q)$ (or choose not to conjugate it) without changing the group of stabilizers.
Similar to before, we can define a set of plaquettes ${\cal P}'_i$ for each $i$, such that
 every plaquette $p$ in ${\cal P}'_i$ has ${\rm dist}(p,i)\leq g/4$ and such that every plaquette $p$ in ${\cal P}'$ with ${\rm dist}(p,i) \leq R$ is in ${\cal P}'_i$ such that the state is stabilized by the set of $A_s$ and by the set of operators
\be
\label{stabsetfinal2}
\Bigl(\prod_{p\in {\cal P}'_i} D^x(p) \Bigr)
\tilde O_i
\Bigl(\prod_{p\in {\cal P}'_i} D^x(p) \Bigr)
\ee
for ${\rm dist}(q,i)>R$.
  For each $i$ such that ${\rm dist}(q,i) \leq g/8$, take ${\cal P}'_i$ to be the set of all plaquettes $p \in {\cal P}'$ within distance $R+g/8$ of $q$.

Now take this set of stabilizers Eq.~(\ref{stabsetfinal2}) and conjugate all stabilizers by $\prod_{p \in {\cal P}'_j} D^x(p)$ for $j$ chosen to be any $1$-cell with ${\rm dist}(q,j)\leq g/8$ (note that all such $1$-cells give the same set ${\cal P}'_j$).  This new set of stabilizers will stabilize a {\it different} state from the state stabilized by Eq.~(\ref{stabsetfinal2}): it will stabilize that state multiplied by  $\prod_{p \in {\cal P}'_j} D^x(p)$.
The resulting state is stabilized by the set of $A_s$ and by
$\tilde O_i$ for $R<{\rm dist}(q,i)\leq g/8$ and also by various other stabilizers.  These other stabilizers are obtained by conjugating the stabilizers for ${\rm dist}(q,i)>g/8$.  We do not bother writing them down, but note the important fact that the support of those stabilizers for any given $i$ is the same as the support of $\tilde O_i$.
Take all these stabilizers and conjugate them by $U_{circuit}^\dagger$; since $U_{circuit}$ need not be Hermitian, let us specify that by ``conjugate" we mean left multiply by $U_{circuit}^\dagger$ and right multiply by $U_{circuit}$.  The resulting state is stabilized by operators $U_{circuit}^\dagger A_s U_{circuit}$, by operators $O_i$ for $R<{\rm dist}(q,i)\leq g/8$, and by various other operators whose support is a distance at least $g/8-2R$ away from $q$.
Now, if we could show that a state existed that was stabilized by the same operators and also by the operators $O_i$ for ${\rm dist}(q,i)\leq R$ then we would be done: we simply undo the conjugation of the stabilizers by $U_{circuit}^\dagger$ and by $\prod_{p \in {\cal P}'_j} D^x(p)$ and we have a state that is stabilizers by the desired set of stabilizers.
However, we claim that such a state does exist: first, we begin by removing some of the stabilizers: we require that the state be stabilized by  $U_{circuit}^\dagger A_s U_{circuit}$ only for ${\rm dist}(s,q)\geq g/16$, while continuing to require that it also be stabilized by operators $O_i$ for $R<{\rm dist}(q,i)\leq g/8$, and by various other operators whose support is a distance at least $g/8-2R$ away from $q$.
Note now that all the operators $U_{circuit}^\dagger A_s U_{circuit}$ have support at least a distance $g/16-R\geq g/32$ away from $q$.
Therefore,
 the only stabilizers which have support within $g/32$ of $q$ are the operators $O_i$.  Each of these operators acts on a single $1$-cell.  Thus, no stabilizer has support on the $1$-cells $i$ for ${\rm dist}(i,q)\leq R$.  Thus, we can also require that the state be stabilized by the operators $O_i$
for ${\rm dist}(q,i)\leq R$.  Thus, we have a state stabilized by
$U_{circuit}^\dagger A_s U_{circuit}$ only for ${\rm dist}(s,q)\geq g/16$, and stabilized by $O_i$ for ${\rm dist}(q,i)\leq g/8$, and by various other operators whose support is a distance at least $g/8-2R$ away from $q$.  Undo the conjugation by $U_{circuit}^\dagger$ and by $\prod_{p \in {\cal P}'_j} D^x(p)$.  This gives a state stabilized by $A_s$ for ${\rm dist}(s,q) \geq g/16$ and also conjugated by the set of operators in Eq.~(\ref{stabset6}) for all $i$.  We simply need to show that the state is also stabilized by the $A_s$ for ${\rm dist}(s,q)<g/16$.  However, note that being stabilized by the set of operators in Eq.~(\ref{stabset6}) for $i$ with ${\rm dist}(i,q)\leq g/16+R+1$ uniquely specifies the reduced density matrix of the state on the set of $1$-cells within distance $g/16+1$ of $q$ (this follows because the toric code obeys the conditions called TQO-1,2 in Refs.~\onlinecite{tqo1,tqo2})
and hence on the set of $1$-cells attached to all $0$-cells within distance $g/16$ of $q$.  Further, we know that a state does exist which is stabilized by the set of operators in Eq.~(\ref{stabset6}) for $i$ with ${\rm dist}(i,q)\leq g/16+R+1$, as such a state can be obtained by unitarily conjugating $\psi$ by $(\prod_{p\in {\cal P}', {\rm dist}(p,i) \leq g/4} D^x(p) )$, and this state is stabilized by $A_s$ for all $s$ within distance $g/16$ of $q$.  Hence, the state that we have constructed is indeed stabilized by all the $A_s$.
\end{proof}
\end{lemma}

We now show that there are no low energy trivial states with no vertex defects.
\begin{theorem}
\label{noletriv}
There is an $\epsilon>0$ such that for all $R<\infty$, for all sufficiently large graphs $G$ in the family of graphs defined above there is no $R$-trivial state $\psi$ on $G \times G$ such that $A_s \psi=\psi$ for all $\psi$ and $|\psi|=1$ and such that
\be
\langle \psi, B \psi \rangle< -(1-\epsilon) N_p,
\ee
where $N_p$ is the number of plaquettes.
\begin{proof}
Pick a set ${\cal P}$ of plaquettes which will soon be treated as a variable.  For given $R$, for sufficiently large $G$, define the state $\phi$ as in lemma \ref{phidefn}.  Define $B'$ by
\be
B'=-\sum_{p \not \in {\cal P}} B_p + \sum_{p \in {\cal P}} B_p.
\ee
That is, $B'$ is also a sum of plaquette terms, just like $B$ is, except the sign of the terms is flipped for plaquettes in ${\cal P}$.
Note that by Eq.~(\ref{local}),
\be
\label{equal}
\langle \psi,B \psi \rangle=\langle \phi,B' \phi \rangle.
\ee
However, we claim that since $b_2(G\times G)$ is extensive, there is an $\epsilon>0$ such that for all sufficiently large $G$ we can find a set of plaquettes ${\cal P}$ such that the minimum eigenvalue of $B'$ is greater than or equal to $-(1-\epsilon)N_p$.  Once we have proven this claim, this will complete the proof, since it will imply, by Eq.~(\ref{equal}) that
\be
\langle \psi,B \psi \rangle\geq -(1-\epsilon) N_p
\ee
for all sufficiently large $G$.

To prove the claim on the minimum eigenvalue of $B'$, note that $B'$ is diagonal in the $S^z$ basis so every eigenvector of $B'$ can be chosen to be an eigenvector of all the operators $S^z_e$ simultaneously.  Let $N_e$ be the number of $1$-cells.  Then, there are $2^{N_e}$ different states which are eigenvectors of all the $S^z_e$ separately: these states correspond to the $2^{N_e}$ different assignments of $+1$ or $-1$ to each $1$-cell.  These states form an orthonormal basis.

Each such basis state $\xi$ is an eigenstate of all the $B_p$.  Number the plaquettes $p$ from $1...N_p$.  So, to each such basis state $\xi$ we can define a vector $v_\xi$ with $N_p$ different entries, with the $p$-th entry of the vector being $+1$ or $-1$ depending upon whether the expectation value of $B_p$ is $+1$ or $-1$.
Each choice of a set ${\cal P}$ can be used to define another vector $v_{\cal P}$: we define this vector to be $-1$ on the $p$-th plaquette if $p \in {\cal P}$ and $+1$ otherwise.  Then, the expectation value
\be
\label{Hamming}
\langle \xi,B' \xi \rangle=-N_p+\sum_p |v_\xi - v_{\cal P}|.
\ee
The term on the right-hand side of the above equation can be understood also in terms of a Hamming distance.  Given a vector with entries $+1$ or $-1$ we can replace each $+1$ with a $0$ and each $-1$ with a $1$ and the result is a bit string.  Then, the term on the right-hand side of the above equation is equal to twice the Hamming distance between the corresponding bit string.  For a given basis state $\xi$, let $s_\xi$ be the corresponding bit string; i.e., the bit string obtained from $v_\xi$ by replacing $+1$ with $0$ and $-1$ with $1$ and let $s_{\cal P}$ be the bit string obtained from $v_{\cal P}$ by a similar replacement.

There are $2^{N_p}$ possible vectors $v_{\cal P}$.  On the other hand, the number of vectors $v_\xi$ is smaller than this by an exponential of an extensive quantity.  Clearly, there are at most $2^{N_e}$ such vectors $v_\xi$ since there are only $2^{N_e}$ basis vectors, but in fact the number of vectors $v_\xi$ is even smaller than this since the mapping from basis states $\xi$ to vectors $v_{\xi}$ is not one-to-one and the number of vectors $v_\xi$ is actually only $2^{N_p-b_2}$, where $b_2$ is the second Betti number.
Since $b_2>0$, there certainly are choices of ${\cal P}$ such that the distance $|v_\xi - v_{\cal P}|$ is positive; i.e., choices of ${\cal P}$ such that the minimum eigenvalue of $B'$ is greater than $-N_p$.  However, we need a stronger result than this.  This stronger result follows by a similar counting argument.
For every bit string of length $N_p$, the number of bit strings within Hamming distance $(\epsilon/2) N_p$ of the given bit string is roughly (up to subexponential corrections)
\be
\exp(N_p H(\epsilon/2)),
\ee
were $H(x)$ is the entropy function: $H(x)=-x \ln(x) - (1-x) \ln(1-x)$.
Thus, the number of bit strings within Hamming distance $(\epsilon/2) N_p$ of the set of strings $s_\xi$ is at most
(up to subexponential corrections) $\exp(N_p H(\epsilon/2)) 2^{N_p-b_2}$.
For sufficiently small $\epsilon$, since $b_2$ is extensive, we can ensure that $\exp(N_p H(\epsilon/2)) < 2^{b_2}$.  Thus, for sufficiently small $\epsilon$, there is some ${\cal P}$ such that $s_{\cal P}$ has Hamming distance greater than $\epsilon/2$ from all $s_\xi$.
Thus, for this ${\cal P}$,
 for all basis states $\xi$ the expectation value $\langle \xi,B' \xi \rangle$  is greater than $-(1-\epsilon) N_p$.
\end{proof}
\end{theorem}

\subsection{An Idea for An Alternate Proof Using Expansion Properties}
\label{cobound}
Let $\delta$ be the coboundary operator.
Suppose the following conjecture holds: ``efficient $\delta^{-1}$"
\begin{conjecture}
There is a constant $c>0$ such that the following holds for all sufficiently large graphs $G$ form the family considered above.
Let $x$ be any $1$-chain with $Z_2$ coefficients on $G \times G$.  Then, there is a $1$-chain $y$ which is a cocycle such that
that
\be
\frac{|{\rm supp}(\delta x)|}{|{\rm supp}(x+y)|} \geq c,
\ee
where $\delta$ is the coboundary operator, ${\rm supp}(...)$ is the support of the coboundary of a vector and $|{\rm supp}(...)|$ denotes the cardinality of the support of a vector.
\end{conjecture}

This conjecture is similar to a cohomological expansion definition in Ref.~\onlinecite{otherexpanders2} but it is crucially different in that we allow $y$ to be any cocycle rather than just a coboundary.  Suppose this conjecture holds.  Under this assumption, we sketch an alternative proof of theorem \ref{noletriv}.  Since the conjecture itself is unproven, we will given only a sketch and many details will be left out.

To motivate the alternate argument, suppose first that we had a state $\psi$ such that $B_p \psi=\psi$ for all $p$.  Then, there would be a certain kind of long-range order in the state $\psi$.  Consider a ``Wilson loop" operator, $W(C)$, which we define to be the product of $S^z_e$ around a $1$-cycle $C$.  If the $1$-cycle is contractible, then $\langle \psi,W(C) \psi \rangle=+1$.  Consider instead cycles $C$ which are the product of a given vertex $v$ in the first graph with a cycle $c$ in the second graph.  We will write $C(v,c)$ to denote such a cycle, determined by the given vertex $v$ and cycle $c$.
Then,
\be
\label{lroWilson}
\langle \psi, W(v,c) \psi \rangle = \langle \psi, W(v',c) \psi \rangle
\ee
for any pair of vertices $v,v'$ and any cycle $c$.
The first step of our sketch is to show that any state with low energy the plaquette terms has a similar (but weaker) kind of long-range order, defined below.

Let $x$ be a $1$-chain.
Consider first a state $\psi_{cl}(x)$ which is a product state, which is an eigenstate of all operators $S^z_e$, with $S^z_e$ having expectation value $-1$ for $e$ in the chain and $+1$ otherwise.  Here the subscript $cl$ stands for ``classical"; we refer later to these states as classical states and they form an orthonormal basis.  
Suppose that $\langle \psi_{cl}(x),-B \psi_{cl}(x) \rangle \leq -(1-\epsilon) N_p$.
Then, by the conjecture above, we can find a $y$ that is a cocycle such that $x+y$ has a small support.  That is $|{\rm supp}(x+y)|\leq \epsilon N_p/c$.
Consider the state $\psi_{cl}(y)$.  This state obeys $B_p \psi_{cl}(y)=\psi_{cl}(y)$ for all $p$, so it has long-range order in the Wilson loop operators as in Eq.~(\ref{lroWilson}).

We can restate the long-range order in Wilson loops in a slightly different way.  Let the graph $G$ have $N$ edges and degree $d$.  Hence the graph has first Betti number $b_1(G)=\frac{d-2}{2}+1$.  For each vertex $v$ in $G$, we define a vector $s(v)$ with $\frac{d-2}{2}+1$ entries; these entries are the expectation values $\langle \psi_{cl}(x), W(v,c) \psi_{cl}(x) \rangle$ for a set of cycles $c$ that form a basis for the first homology.
The entries in this vector equal $\pm 1$.

Note that if we instead consider a vector $t(v)$ with entries being the expectation values  $\rangle \psi_{cl}(y), W(v,c) \psi_{cl}(y) \rangle$, then
$t(v)=t(v')$ for all $v,v'$.  The chain $x+y$ has small support.  Let $N(x+y,v)$ denote the number of $1$-cells in $x+y$ which are a product of the given vertex $v$ in the first graph and an arbitrary edge in the second graph.  Note that $\sum_e N(x+y,v)\leq |{\rm supp}(x+y)|$.  So, speaking loosely, for most $v$, we have $N(x+y,v)$ small compared to $N$.  For example, for at least $1/2$ of $v$, we have
\be
N(x+y) \leq 2 \frac{\epsilon}{c} \frac{N_p}{N}=2 \frac{\epsilon}{c}\frac{N_p}{N^2} N.
\ee
Note that $N_p/N^2=O(1)$, so for at least $1/2$ of $v$, $N(x+y,v)$ is bounded by $O(1) \epsilon N=O(1) \epsilon N_e$.  (The specific number $1/2$ is not important.)

We now define a distance function and use this bound on $N(x+y,v)$ to bound the distance between $s(v)$ and $s(v')$.
Note that the vector $s(v)$ is completely determined by the vector of expectation values $\langle \psi_{cl}, S^z_e \psi_{cl} \rangle$ for $1$-cells $e$ which are a product of the given vertex $v$ in the first graph and an arbitrary edge in the second graph.  This second vector has $N_e$ entries.  In fact, there are many vectors of expectation values $\langle \psi_{cl}, S^z_e \psi_{cl} \rangle$ which determine the same vector $s(v)$.
In general, we will say that ``$s(v)$ is the Wilson loop vector of $w$" if the entry of $s(v)$ corresponding to cycle $c$ is equal to the product of the entries of $w$ corresponding to the edges in that cycle.
We define a distance function between two vectors, $s(v)$ and $s(v')$ as follows.   Take the minimum, over all vectors $w,w'$ such that $s(v)$ is the Wilson loop vector of $w$ and $s(v')$ is the Wilson loop vector of $w'$, of the Hamming distance between $w$ and $w'$.

Then, we have the for at least $1/2$ of $v$, the distance between $s(v)$ and $t(v)$ is at most $O(1) \epsilon N_e$.
Then, for at least $1/4$ of the pairs $v,v'$ the distance between $s(v)$ and $s(v')$ is at most $2 O(1) \epsilon N_e=O(1) \epsilon N_e$.
The number of possible choices of at most $O(1) \epsilon N_e$ edges from a set of $N_e$ edges is $\exp(O(1) \epsilon N_e)$, and for at least $1/4$ of the pairs $v,v'$, given $s(v)$ the vector $s(v')$ is one of at most $\exp(O(1) \epsilon N_e)$ possible vectors.

We now define a quantum operator that we write as $\delta_{W(v),z)}$.  The notation is intended to be suggestive here: $z$ is an arbitrary vector with $b_1(G)$ entries, the entries equaling $\pm 1$, and this operator will compute the probability that the vector of Wilson loop expectation value is equal to $z$.
We define
\be
\delta_{W(v),z} = \prod_c \frac{1+W(v,c) z_c}{2}
\ee
where the product is over
a set of cycles $c$ that form a basis for the first homology and $z_c$ denotes the entry of $z$ correspnding to cycle $c$.  The operators 
\be
\frac{1+W(v,c) z_c}{2}
\ee
are projectors which commute with each other, so $\delta_{W(v),z}$ is a projector.
Note that
\be
\langle \psi_{cl}(x), \delta_{W(v),z} \psi_{cl}(x) \rangle 
\ee
equal $1$ if $z=s(v)$ and $0$ otherwise.

Compute the sum of expectation values
\be
\sum_{z,z'}^{{\rm dist}(z,z')\leq O(1) \epsilon N_e}
 \langle \psi_{cl}(x), \delta_{W(v),z} \delta_{s(v'),z'} \psi_{cl}(x) \rangle
\ee
where the sum is over vectors $z,z'$ with $b_1(G)$ entries each having values $\pm 1$.
For randomly chosen $v,v'$, this expectation value is at least $1/4$.  
This is the ``long-range order" that we identify for this state $\psi_{cl}$.  We now consider any arbitrary state $\psi$ (not necessarily
a classical state) such that $\langle \psi,-B \psi \rangle \leq -(1-\epsilon/2) N_p$.  This state $\psi$ is a superposition of classical states: $\psi=\sum_{x} A(x) \psi_{cl}(x)$.  Choosing a classical state $\psi_{cl}(x)$ with probability distribution $|A(x)|^2$, with probability at least $1/2$ we find that 
$\langle \psi_{cl}(x),-B \psi_{cl}(x) \rangle \leq -(1-\epsilon) N_p$
Thus, for any such state $\psi$, we find that the expecation value
\be
\sum_{z,z'}^{{\rm dist}(z,z')\leq O(1) \epsilon N_e}
 \langle \psi, \delta_{W(v),z} \delta_{s(v'),z'} \psi \rangle
\ee
is at least $1/8$.

Note, however, that if $\psi$ is an $R$-trivial state, then the expectation value factors if the distance between $v$ and $v'$ is greater than $2R$ (note that here by distance we are referring to the distance on the graph, using the graph metric).  Further, for a randomly chosen pair $v,v'$, the probability that the distance between $v,v'$ is greater than $2R$ tends to $1$ as $N \rightarrow \infty$.
Thus, up to corrections that go to zero as $N \rightarrow \infty$, we find that 
\be
\label{factor}
\sum_{z,z'}^{{\rm dist}(z,z')\leq O(1) \epsilon N_e}
 \langle \psi, \delta_{W(v),z} \delta_{s(v'),z'} \psi \rangle=
\sum_{z,z'}^{{\rm dist}(z,z')\leq O(1) \epsilon N_e}
 \langle \psi, \delta_{W(v),z} \psi \rangle \langle \psi, \delta_{s(v'),z'} \psi \rangle.
\ee

Finally, suppose that $\psi$ had the property that for all $v$ we had that 
\be
\langle \psi, \delta_{W(v),z} \psi = 2^{-b_1(G)}.
\ee
That is, suppose that there was a uniform probability distribution, independent of $z$, of having different Wilson loop vectors.
This can be seen to contradict Eq.~(\ref{factor}) for sufficiently  small $\epsilon$, since in that case the fraction
\be
 \frac{\exp(O(1) \epsilon N_e)}{2^{b_1(G)}}
\ee
tends to zero as $N_e\rightarrow\infty$ for small enough $\epsilon$ and this fraction represents the fraction of vectors within distance $O(1)\epsilon N_e$ of a given vector.
Hence, we cannot have a uniform probability distribution.

However, it can be shown that having a non-uniform probability distribution contradicts the assumption of a trivial state with no vertex defects.
The argument involves constructing operators similar to the $C^x(p)$ and $D^x(p)$ above.  For any $1$-cell $e$ in $G \times G$ which is a product of a vertex $v$ in the first graph and an edge $a$ in the second graph, we can construct an operator $O$ that anti-commutes with $S^z_e$ and which commutes with all other $S^z_f$ for $f$ being a product of vertex $v$ with an edge $b \neq a$.  If the girth of the graph is large enough compared to $R$, we can construct this operator $O$ such that it commutes with all the plaquette operators $B_p$ except for certain plaquttes $p$ far from $v\times G$ (note: this means that the plaquettes are far from all $1$-cells $e$ of the form $v$ times an edge in the graph, not just far from $e$); this operator $O$ will anti-commute with $B_p$ for $p$ in some set $S$ of plaquettes with ${\rm dist}(S,v \times G)>2R$ and will commute with all other $B_p$.  Just like the $C^x(p)$ and $D^x(p)$ were, this operator $O$ will be a product of $S^x_f$ over $f$ in some set of $1$-cells.
We can also construct an operator $O'$ such that $O O'$ is a product of operators $A_s$.  Then,
$O O' \psi = \psi$.  
This argument that we now follow is quite similar to that in section VIII of Ref.~\onlinecite{leshouches}.
 Further, we can choose $O,O'$ such that the intersection of the support of $O$ with the set of sites within distance $2R$ of $v\times G$ is at least distance $2R$ away from the intersection of the support of $O'$ with the set of sites within distance $2R$ of $v \times G$.
Using the assumption that the state is $R$-trivial, we can show that $\psi$ and $O \psi$ have the same reduced density matrices on $v \times G$ (to see this, pull back on the quantum circuit, so that we consider operators $\tilde O=U_{circuit}^\dagger O U_{circuit}$ and $\tilde O'=U_{circuit}^\dagger O' U_{circuit}$ acting on a product state ; using the distance between the supports of $O$ and $O'$ mentioned above, we can see that on the set of $1$-cells in $v\times G$), the supports of $\tilde O$ do not overlap; since $\tilde O \tilde O'$ acting on the product state gives the product state, this implies that $\tilde O$ leaves the reduced density matrix on $v\times G$ unchanged).
So,
\be
\langle \psi, \delta_{W(v),z} \psi \rangle = \langle \psi, O \delta_{W(v),z)} O \psi \rangle=\langle \psi,\delta_{W(v),z'} \psi \rangle,
\ee
where $z'$ is the vector obtained by finding a $w$ such that $z$ is the Wilson loop vector of $w$, then flipping the value of the entry in $w$ corresponding to edge $a$ to obtain a vector $w'$ and then letting $z'$ be the Wilson loop vector of $w'$.
Using the above equation, for various choices of $a$, we can see that $\langle \psi, \delta_{W(v),z} \psi \rangle$ indeed must be independent of $z$ for this $R$-trivial state.

\subsection{Triviality of Low Energy States with Vertex Defects}
Theorem \ref{noletriv} showed the absence of trivial low energy states without any vertex defects.  Now, we show the converse: there are trivial low energy states with vertex defects.  Indeed, there are trivial states with any arbitrarily low density of vertex defects and with no plaquette defects.
\begin{theorem}
\label{leverttriv}
For any $\epsilon>0$, there exists an $R$ such that  for all large sufficiently graphs $G$ in the family of graphs defined above there is an $R$-trivial state $\psi$ on $G \times G$ such that $B_p \psi=\psi$ for all $\psi$ and such that
\be
\langle \psi,A \psi \rangle \leq -(1-\epsilon) N_v,
\ee
where $N_v$ is the number of vertices.
\begin{proof}
We construct the state $\psi$ by picking a subset ${\cal S}$ of the vertices, with the subset containing a fraction $\epsilon$ of the vertices, and construct an $R$-trivial state that is an exact ground state of the Hamiltonian
\be
H_{tc}({\cal S})=B-\sum_{s\not\in {\cal S}} A_s.
\ee
 That is, it is an exact ground state of the Hamiltonian with the operators $A_s$ for $s\in {\cal S}$ removed.

We construct this $R$-trivial state by giving a unitary quantum circuit that turns $H_{tc}({\cal S})$ into a Hamtilonian that is diagonal in the $S^z$ basis and hence which clearly has a product state ground state.  The procedure that follows is closely related to that in Appendix A of Ref.~\onlinecite{mixedtriv}.
Consider a given vertex $s\in {\cal S}$.  Consider any $1$-cell $e$ attached to this vertex.  Note that there are two $0$-cells attached to $e$ and so there is some other $0$-cell $s'$ attached to $e$ with $s' \neq s$.  Assume that $s' \not \in {\cal S}$.  Define the unitary transformation
\be
U_{e,s'}=\exp(\frac{\pi}{4} S^z_e A_{s'}).
\ee
Then,
\be
U_{e,s} H_{tc}({\cal S}) U_{e,s'}^\dagger=H_{tc}({\cal S} \cup \{s'\})-S^z_e.
\ee
Define a unitary $U_{s}$ to be the product of $U_{e,s'(e,s)}$ over a subset (given in the next sentence) of the $1$-cells $e$ attached to $s$ with $s'(e,s)$ being the $0$-cell other than $s$ that is attached to $e$.  The particular subset we choose is the subset of $e$ such that $s'(e,s) \not \in {\cal S}$.
\be
U_s H_{tc}({\cal S}) U_s^\dagger = H_{tc}({\cal S}'),
\ee
where ${\cal S}'$ is equal to the union of ${\cal S}$ with the set of all $0$-cells attached to a $1$-cell attached to $s$.

We now define the circuit.  Let ${\cal S}_0={\cal S}$.  Let $H_0=H({\cal S})$.  Let $H_i=U_i H_{i-1} U_i^\dagger$.  We will choose the unitaries $U_i$ so that each Hamiltonain $H_i$ is equal to $H_{tc}({\cal S}_i)$ plus some sum of operators $S^z_e$, with the sum being taken over a set of edges $E_i$.  The set of edges $E_i$ will have the property that every edge in $E_i$ is not in the support of any operator in the sum $-\sum_{s\not\in {\cal S}} A_s$, and hence the Hamiltonian will still be a set of commuting terms.

For given $\epsilon$, we pick ${\cal S}$ to be a set of $s$ such that every $s'$ is within some bounded ($\epsilon$-dependent) distance of ${\cal S}_0$.
The $i$-th round of the circuit will be a product of $U_s$ over some given set $X_i$ of $s$.  We choose a set $X_i$ such that there is no pair $s,s'\in X$ with ${\rm dist}(s,s') \leq 2$.  We claim that we can choose such sets $X_i$ such that after a bounded number of rounds, all $0$-cells are in ${\cal S}$, leaving a Hamiltonian that is diagonal in the $S^z$ basis.  Here is one way to show this claim: color the set of all $0$-cells by some bounded number of different colors such that no two $0$-cells within distance two of each other have the same color.  This can be done with $c=O(1)$ colors.  Let the $i$-th round of the circuit have $X_i$ all $i$ with color $i \, {\rm mod} \, c$ such that $s \in {\cal S}_i$.  Then, after $c$ rounds, ${\cal S}_c$ will contain all $s$ within distance $1$ of ${\cal S}_0$, after $2c$ rounds, ${\cal S}_{2c}$ will contain all $s$ within distance $2$ of ${\cal S}_0$, and so on, and so after a bounded number $k$ of rounds, ${\cal S}_k$ will contain all $s$.
\end{proof}
\end{theorem}

\section{Hypergraph Codes}
\label{hypgc}
This above section naturally leads to the question of whether we can construct a Hamiltonian which has NLTS, without making this ``one-sided" assumption.  One interesting way to attack this question is to consider the hypergraph product codes of Ref.~\onlinecite{hpc}.
 We will use a slightly different language, the language of chain complexes, to define these codes.  This language will make certain generalizations possible, as we discuss in the next two subsections.  Ref.~\onlinecite{hpc} uses the language of hypergraphs; we begin by describing hypergraphs as chain complexes.

\subsection{Hypergraphs and Hypercomplexes}
We can define a hypergraph with hypervertices and hyperedges by a bipartite graph.  The graph is divided into left and right parts, with the left part of the graph representing the hyperedges of the hypergraph and the right part of the graph representing the hypervertices, with an edge in the graph between a right node and a left node if the hypervertex corresponding to the right node is in the hyperedge corresponding to the left node.
Given a graph representation of a hypergraph, we can define a space $C_1$ (this space may be a vector space or it may be more general; see below) with dimensionality equal to the number of left nodes and a space $C_0$ with dimensionality equal to the number of right nodes, and a linear operator $\partial_1$ called the ``boundary operator" from $C_1$ to $C_0$, with the linear operator $\partial_1$ being the adjacency matrix of the graph.
This representation suggests a natural generalization of hypergraphs to ``hypercomplexes".  Such a hypercomplex in fact is simply a chain complex
with certain sparseness restrictions on the boundary operators.

A chain complex $C$ is defined by a sequence of spaces $C_i$ and by a sequence of maps $\partial_i$, which we call boundary operators, where $\partial_i$ maps from $C_i$ to $C_{i-1}$ such that $\partial_{i-1} \partial_i=0$.  Each $C_i$ will be taken to be either $\zc^{D_i}$ for some given integer $\coeff$ and some given integers $D_i$, or more generally we can take $C_i=R^{D_i}$ for any ring $R$.
Then, the boundary operator $\partial_i$ can be written as a matrix with $D_{i-1}$ rows and $D_i$ columns , with entries of the matrix in the given ring.
We will be interested in the case that the boundary operator is sparse: each row should have $O(1)$ non-zero entries in it and each column should also have $O(1)$ non-zero entries.  Note that in the case of a boundary operator corresponding to a simplicial complex or cubical complex or other polyhedral complex, the operator $\partial_1$ has two non-zero entries in every column (since every $1$-cell has two $0$-cells in its boundary) and the number of non-zero entries in a given row is equal to the number of $1$-cells attached to a given $0$-cells and hence is bounded if the local geometry of the complex bounded.
Note that if $\coeff$ is prime, $\zc \cong \fc$ so the $C_i$ are vector spaces with dimensionality $D_i$.

It is possible to take products of such hypercomplexes.  The product we now describe is known in the math literature as constructing a bicomplex from two chain complexes, and then taking the total complex of that chain complex; one important point for us is that that process will preserve the sparsity properties of the boundary operators.
Given two complexes, one complex $C$ given by a sequence $C_0,C_1,...$ with boundary operators $\partial_i$, and the other complex $C'$ given by a sequence $C'_0,C'_1,...$ with boundary operators $\partial'_i$, we define the product of the two complexes $C \times C'$ by the sequence
\be
C_0 \otimes C'_0, \; \Bigl( C_0 \otimes C'_1 \Bigr) \oplus \Bigl(C_1 \otimes C'_0 \Bigr), \; ...
\ee
That is, the $j$-th space in $C \times C'$ is given by
\be
\oplus_{i=0}^j \Bigl( C_i \otimes C'_{j-i} \Bigr).
\ee

Let $\tilde \partial_j$ be the boundary operators for $C \times C'$.
We define these boundary operator $\tilde \partial_j$ by describing its action on vectors in each of the vector spaces $C_i \otimes C'_{j-i}$ for $i=0,...,j$.
If $v_i $ is a vector in $C_i$ and $v'_{j-i}$ is a vector in $C'_{j-i}$ we define
\be
\label{compderivprod}
\tilde \partial_j \Bigl( v_i \otimes v'_{j-i} \Bigr) = (\partial_j v_i)\otimes v'_{j-i} + (-1)^i v_i \otimes \partial'_{j-i} v'_{j-i}.
\ee
The signs $(-1)^i$ are chosen so that
\be
\label{isacomplex}
\tilde \partial_{i-1} \tilde \partial_i=0.
\ee

By using this definition for the product of two complexes, one can define the product of three or more complexes.  The product is associative so that $(C \times C') \times C''= C \times (C' \times C''$, so we simply write products of complexes without parentheses.

\subsection{$\zc$ Quantum Codes Based on Hypercomplexes}
One important aspect of the above definition of hypercomplexes is that Eq.~(\ref{isacomplex}) holds with arbitrary coefficients, including real number coefficients, and not just for $Z_2$ coefficients.  If in Eq.~(\ref{compderivprod}) we had not included the sign $(-1)^i v_i$ then the identity Eq.~(\ref{isacomplex}) would have held with $Z_2$ coefficients but not otherwise.
This will allow us to define $\zc$ codes based on hypercomplexes for arbitrary integers $\coeff$.

In this section, we will generalize the construction of Ref.~\onlinecite{hpc} in two ways.  First, as mentioned, we will extend this construction to $\zc$ codes using our definition of hypercomplexes.  Second, we will generalize to hypercomplexes which are the product of three or more hypergraphs.

Rather than review the construction of Ref.~\onlinecite{hpc}, we re-express the construction in other language.  For the case of a hypercomplex based on the product of two hypergraphs using a $\zc$ code with $\coeff=2$, with the integer $q$ below equal to $1$, this repeats the construction of Ref.~\onlinecite{hpc}.  For other complexes, for $n \neq 2$, or for $q \neq 1$, this is a slightly more general construction.  For example, one could take the product of four hypergraphs to obtain a chain complex with spaces $C_0,...,C_4$ and pick $q=2$ to obtain a different construction.

Pick an integer $q$.
We define a code with a number of degrees of freedom equal to $D_q$.
For each such degree of freedom we have an $\coeff$-dimensional Hilbert space.
We define matrices
$U$ and $V$ as the ``clock" and ``shift" matrices.  $U$ is a diagonal matrix with entries $1,\exp(2 \pi i/\coeff), \exp(4 \pi i/\coeff),...$ and $V$ is a permutation matrix with a $i$ in the $i,j$ position of the matrix if $i=(j+1)\, {\rm mod} \,\coeff$.
We define the Hamiltonian by
\be
H=A+B,
\ee
where
\be
\label{Adef2}
A=-\sum_{k=1}^{D_{q-1}} A_k,
\ee
and
\be
\label{Bdef2}
B=-\sum_{k=1}^{D_{q+1}} B_k.
\ee
In the case that $q=1$, the sums of Eqs.~(\ref{Adef2},\ref{Bdef2}) are analogous to the sum over $0$-cells and $2$-cells used in Eqs.~(\ref{Adef},\ref{Bdef}).  For example, in Eq.~(\ref{Adef2}), we sum over $k=1...D_0$ for $q=1$, and so we have one term in the sum for each $0$-cell.

Let us introduce some notation.  We write matrix elements of the boundary operator $\partial_i$ by $(\partial_i)_{l,m}$, indicating the entry in the $l$-th row and $m$-th column, where we labels the rows by integers $1,...,D_{i-1}$ and the columns by integers $1,...,D_i$.
We define $A_k$ by
\be
A_k=\prod_{l}U_e^{(\partial_q)_{k,l}} + h.c..
\ee
We define $B_k$ by
\be
B_k=\prod_{l}V_e^{(\partial_{q+1})_{l,k}} + h.c..
\ee
Because $\partial_{i-1} \partial_i=0$, we have $[A_k,B_l]=0$ for all $k,l$, for all  chain complexes.

\subsection{Triviality Properties of Different Hypergraph Codes}
The particular Hamiltonian we considered in Section \ref{tconcomplex} on $G \times G$ corresponds to a particular example of codes formed in this manner.   This code does not have a duality between the electric and magnetic degrees of freedom (or rather, the duality that interchanges the electric and magnetic degrees of freedom does not have leave the complex unchanged).
As that imbalance between electric and magnetic degrees of freedom might be responsible for the one-sided nature of our result, it is natural to consider instead a more balanced case.

Suppose we consider a hypergraph which has the same number of hyperedges and hypervertices.  In our language, this hypergraph correspnds to a bipartite graph with the same number of vertices in each part, or equivalently to a chain complex where the boundary operator $\partial_1$ is a square matrix.  Suppose we choose such a hypergraph at random, with each hyperedges containing $d$ hypervertices and each hypervertex in $d$ hyperedges, for some small $d$ such as $d=4$.  Let us fix $\coeff$ to be a prime to make the language simpler: it will enable us to talk about the dimension of various vector spaces (fixing $\coeff$ prime is not essential).
For typical random such hypergraph, the zeroth and first Betti numbers of the complex will be $O(1)$.  As a result, if we define a complex by taking the product of this complex with itself, then the resulting complex will having zeroth, first, and second Betti numbers all being $O(1)$.

The second Betti number counts the ``redundancy" of the operators $B_k$.  That is, for any $v \in C_2$ such that $\partial_2 v=0$,
the operator
\be
\prod_k B_k^{v_k}
\ee
is exactly equal to the identity operator; here we define $v_k$ to be the $k$-th entry of vector $v$.  The second Betti number is the dimension of the space of such vectors because we $C_3=0$.
Similarly, for any vector $v$ in $C_0$ such that $\partial_1^\dagger v=0$,
\be
\prod_k A_k^{v_k}
\ee
equals the identity.  The zeroth Betti number counts the ``redundancy" of the operators $A_k$.  Note that if the complex $C$ had been obtained from a simplicial complex then the zeroth Betti number is always non-zero and the product of all the $A_k$ is equal to the identity in that case.  However, for an arbitrary chain complex $C$ the zeroth Betti number can indeed be zero.

We now follow an argument due to Yoshida\cite{yoshida} (Yoshida applied this argument to the case of Haah's cubic code\cite{haah}).  This argument shows the absence of a phase transition as a function of temperature for a Hamiltonian with subextensive redundnacy of constraints.  Here we give this argument for a Hamiltonian with $q=1$ given a sequence of complexes for which the zeroth and second Betti numbers are both subextensive and for which $C_3=0$.

We first consider the case in which both Betti numbers are zero so that there is no redundancy: that is, there is nontrivial product of the operators $A_k, B_k$ that is equal equal to the identity.
Consider first the case $p=2$, where the groups that we use are more familiar.  Then, there is a qubit for each $e$ and $U_e$ is the spin operator $S^x_e$ and $V_e=S^z_e$.
Then, we can find a unitary $U_{Cliff}$ in the Clifford group to convert the Hamiltonian into the following canonical form:
\begin{eqnarray}
U_{Cliff} A_k U_{Cliff}^\dagger &=& 2 S^x_k, \\ \nonumber
U_{Cliff} B_k U_{Cliff}^\dagger &=& 2 S^x_{k+D_0}.
\end{eqnarray}
Note that the integer $k$ labelling the different $A_k$ ranges from $1,...,D_0$, while the label $e$ in $S^x_e$ ranges from $1,...,D_1$.  Thus, we map the $D_0$ different operators $A_k$ into the operators $2 S^x_k$ for $k=1,...,D_0$ and we map the $D_2$ different operators $B_k$ into the operators $2 S^x_{k+D_0}$, wth $k+D_0$ ranging from $D_0+1,...,D_0+D_2$.  Note that given that both Betti numbers are zero, we have $D_1 \geq D_0+D_2$.
The factor of $2$ appears on the right-hand side of the above equation because we have defined each operator $A_k,B_k$ to equal a product of Pauli operators plus its hermitiian conjugate, hence each $A_k,B_k$ is equal to two times a product of Pauli operators.
The existence of such a unitary $U_{Cliff}$ follows from the absence of any redundancy of the operators, and can be constructed using a process similar to Gaussian elimination as follows.  Define a rectangular matrix with entries in $F_2$, with $D_0$ rows and $D_1$ columns, with a $1$ in a given matrix element $k,l$ if the operator $A_k$ for $k$ corresponding to the given row contains the operator $S^x_l$ for $l$ corresponding to the given column.  
For any $i,j$ we can find an unitary $W$ in the Clifford group such that $W S^x_i W^\dagger=S^x_i S^x_j$ and such that $W S^x_k W^\dagger=S^x_k$ for $i \neq k$.  (This unitary $W$ is simply a controlled-NOT gate in the $x$-basis).  So, conjugating with this matrix $W$ changes the entries of the matrix above by adding one column to another.  Then, by a process of Gaussian elimination, we can use this column addition to bring the matrix into the form that matrix element in position $k,l$ is equal to $\delta_{k,l}$.  Here we use the assumption of no redundancies to show that the original matrix is non-singular.  The unitary corresponding to this sequence of row additions changes the operators $B_k$ in some way; since the $A_k,B_k$ commute, the operators $B_k$, after conjugation by this unitary, are supported only on sites $D_0+1,...D_1$.  Then, we can apply a Hadamard transformation to make the $B_k$ diagonal in the $x$-basis and then apply a similar Gaussian elimination to bring them into the canonical form too.

For $p>2$, the operators $U,V$ generate an extra-special group, rather than the Pauli group.  The group of normalizers of this extra-special group in the unitary group can be regarded as a generalization of the Clifford group\cite{genCliff}.  We can find a unitary in this group of normalizers to make a similar transformation to the equaton above, bringing the operators into the canonical form:
\begin{eqnarray}
U_{norm} A_k U_{norm}^\dagger &=& U_k + h.c., \\ \nonumber
U_{norm} B_k U_{norm}^\dagger &=& U_{k+D_0} + h.c.
\end{eqnarray}
Define the partition function to be
\be
Z(\beta)={\rm tr}\Bigl( \exp(-\beta H) \Bigr).
\ee
Define the thermal expectation value of the energy to be
\be
E(\beta)=Z(\beta)^{-1} {\rm tr}\Bigl( H \exp(-\beta H) \Bigr).
\ee
Note that $Z(\beta)$ and $E(\beta)$ are both invariant under replacing $H \rightarrow U_{norm} H U_{norm}^\dagger$, since $U_{norm}$ is unitary.  However, after the operators $A_k,B_k$ are brought into the given canonical form, the calculation of the partition function and energy becomes trivial, since everything manifestly decouples: for different $k$, the operators $U_k$ have different support, and we have to solve $D_0+D_2$ different problems, each problem  on a single site.
So we can exactly calculate $E(\beta)$ for this problem.  
Let
\be
E^A_k(\beta)=\frac{{\rm tr}\Bigl(A_k \exp(-\beta A_k) \Bigr)}{{\rm tr}\Bigl(\exp(-\beta A_k) \Bigr)},
\ee
and let
\be
E^B_k(\beta)=\frac{{\rm tr}\Bigl(B_k \exp(-\beta B_k) \Bigr)}{{\rm tr}\Bigl(\exp(-\beta B_k) \Bigr)},
\ee
Then,
\be
Z(\beta)^{-1} {\rm tr}\Bigl( A_k \exp(-\beta H) \Bigr)=E^A_k(\beta)
\ee
and
\be
Z(\beta)^{-1} {\rm tr}\Bigl( B_k \exp(-\beta H) \Bigr)=E^B_k(\beta).
\ee
That is, the thermal expectation value of $A_k$ is the same whether or not the other terms $A_l$ for $l\neq k$ and $B_l$ are present in the Hamiltonian,
and similarly for the thermal expectation value of $B_k$.
As a result,
\be
E(\beta)=-\sum_{k=1}^{D_{q-1}} E^A_k(\beta)-\sum_{k=1}^{D_{q+1}} E^B_k(\beta).
\ee
The functions $E^A_k,E^B_k$ can be calculated easily.  In fact they are all equal to each other, but that fact is not important.  In the case $\coeff=2$, they are simply hyperbolic tangent functions.  The crucial thing is that 
these functions $E^A_k(\beta),E^B_k(\beta)$ are all analytic functions and are independent of the particular choice of the complex, and so if we consider a family of complexes then the thermal expectation value of the energy density
\be
\frac{E(\beta)}{D_{q-1}+D_{q+1}}
\ee
is an analytic function and so there is no thermal phase transition as a function of $\beta$.

Even if the zeroth and second Betti numbers are non-zero, if we consider a sequence of complexes in which they are subextensive the limiting thermal expectation value of the energy density will still be an analytic function.  To see this, remove a subextensive number of terms from the Hamiltonian to obtain a new complex with vanishing zeroth and second Betti number, and that complex will have an analytic limiting thermal expectation value of the energy density.  Removing a subextensive number of number of terms will change the energy density by $o(1)$ and so will not change the limiting value.

However, so far as we know, the fact that there is no phase transition in the thermodynamics does not necessarily imply that there is no $\beta_c<\infty$ such that 
for $\beta\geq \beta_c$ the thermal state $Z(\beta)^{-1} \exp(-\beta H)$ is nontrivial in the sense of Ref.~\onlinecite{mixedtriv}.  This is an interesting problem for the future.

Still, this Hamiltonian is an interesting one to consider.  If indeed there is such a $\beta_c$ above which there is a transition to a circuit nontrivial state, then perhaps this Hamiltonian might also have NLTS.
Can we find a better candidate Hamiltonian?  The Hamiltonian of section \ref{tconcomplex} had a lot of redundancy, which we used crucially in the one-sided proof, but it suffered from the fact that the degrees of freedom were on $1$-cells and each $1$-cell was attached to only $2$ $0$-cells.
So, another thought is to replace the unbalanced $G\times G$ case of section \ref{tconcomplex} with the product of a hypergraph with itself, choosing a hypergraph with extensive first Betti number so that the resulting complex will still have lots of redundancy (extensive second Betti number) and such that the resulting complex will have each $1$-cell attached to more than $2$ $0$-cells.

To do even better, a natural guess is that we want a lot of redundancy in both the $A_k$ operators and the $B_k$ operators.  By choosing an extensive second Betti number, we obtain a large redundancy in the $B_k$ operators.  Can we obtain an extensive redundancy in both $A_k$ and $B_k$ operators?
This seems to require going to $q>1$.  In the discussion later we consider whether this is possible; see also Open Question \ref{oq2.10}.

However, homology or cohomology is not the only way to obtain a redundancy in the operators.
Consider a $d$-fold product of a hypergraph chosen to have extensive first Betti number.  This $k$-fold product has extensive $d$-th Betti number.  Take $q=k-1$.  Then, there is an extensive redundancy in the operators $B_k$.  The $d-2$-th Betti number may not be extensive.  However, there can still be redundancy in the operators $A_k$ if $d>2$.  That is, there may still be $v\in C_{d-2}$ such that $v$ is a cocycle.  Indeed, choosing $v$ to be any coboundary suffices.  Thus, a final interesting Hamiltonian to consider is such a $d$-fold product, perhaps for large $d$.

\section{Discussion}
Perhaps the most obvious question we have left open is this: can we construct some families of complexes and Hamiltonians with NLTS?  The Hamiltonians we considered in this paper were toric code Hamiltonians.  One route to constructing the desired family of complexes and Hamiltonians might be to consider other types of Hamiltonians, but let us first consider whether we could achieve this goal using toric code Hamiltonians.
Let us give some intuition on this.  We think of expander graphs as being complexes on which particles can move very easily: a particle is a $0$-dimension object and a path is a $1$-dimensional object and even deleting a sufficiently small nonzero fraction of vertices from an expander graphs gives a graph which has long paths.  We think of a non-$1$-hyperfinite complexes as being a complex on a which a string, a $1$-dimensional object, can move very easily, as the path of a string is a $2$-dimensional object and we can delete a small fraction of cells from a non-$1$-hyperfinite complex and get a complex which has large surfaces.  Theorem \ref{noletriv} considers a toric code Hamiltonian on a $2$-complex that was not a triangulation of a manifold.  However, we believe that a similar result could be proven using a $(1,3)$-code on the $4$-complex which is the product of a hyperbolic surface with itself, a complex which is a triangulation of a manifold and which is non-$1$-hyperfinite.  Such a $(1,3)$-code has degrees of freedom on $1$-cells, and has the same Hamiltonian as $H_{tc}$ above: there are terms in the Hamiltonian $A_s$ for each $0$-cell and $B_p$ for each $2$-cell and
the result that we conjecture would exist in this case is that there is no low energy trivial state in which all the $A_s$ terms are in their ground state (no low energy trivial state without electric defects).
Now, suppose we want to come up with a $(q,r)$-code on a $d$-complex for which there is NLTS without imposing any restriction that there be no electric defects.  A natural guess is that we need the complex to be both non-$q$-hyperfinite and non-$r$-hyperfinite.  Perhaps we want the complex to be a triangulation of a $d$-dimensional hyperbolic manifold.  See Open Question \ref{oq2.10} for the question of whether such a complex exists.  If $d$ is even, the most favorable situation to achieve this would be $q=r=d/2$ while if $d$ is odd the most favorable situation is $q=\lfloor d/2 \rfloor, r=\lceil d/2 \rceil$ or $q=\lceil d/2 \rceil, r=\lfloor d/2 \rfloor$.  So, we want a $d$-complex that is non-$\lceil d/2 \rceil$-hyperfinite.

One motivation to consider complexes which are both non-$q$-hyperfinite and non-$r$-hyperfinite is based on the result in the appendix of Ref.~\onlinecite{mixedtriv}.  There, it was shown that for a $(1,2)$ code on a $3$-dimensional cubic lattice with a small density of holes removed from the lattice there was a trivial ground state.  This lattice with a small density of holes removed is $2$-localizable with a range dependent upon the density of holes This result was based on specific properties of this $(1,2)$ code while for more general Hamiltonians it may not be case that such a trivial state exists.  The result uses the construction of a specific unitary quantum circuit for this code.  It seems that the ideas in this construction could perhaps be extended to show that if the complex is $q$-localizable or $r$-localizable that a $(q,r)$ code would have a trivial ground state and hence that if the complex is $q$-hyperfinite or $r$-hyperfinite that a $(q,r)$ code would have a trivial low energy state.

\appendix

\section{Notions for infinite graphs and noncompact spaces}

Elek\cite{hyperfinite} defined a family of graphs $\{X_i\}$ to be hyperfinite if $\forall$ $\epsilon>0$ $\exists$ $r>0$ such that by deleting a fixed fraction of edges the resulting components each had diameter $<r$.

We have renamed Elek's notion ``$0$-hyperfinite'' and made the

\noindent{\underline{Definition:}}
A family of simplicial complexes (with metrics) $\{X_i\}$ is $k$-hyperfinite if $\forall$ $\epsilon>0$ $\exists$ $r>0$ such that by deleting an $\epsilon$-fraction of the simplicies of each $X_i$ to get $\{Y_i\}$, then there are continuous functions $f_i:Y_i\to K_i$, $K_i$ a $k$-complex so that $\text{diam}(f_i^{-1}(p))<r$ for all $p\in K_i$.

In (arXiv:1105.320) Abert and Elek showed that the group graphs $C$ of an amenable group $G$ (and the Cayley graphs of relatively amenable subgroups) exhibited the hyperfinite property ``all at once'': for all $\epsilon>0$, there is an $r>0$ and an edge set $E_\epsilon$ of density $\leq\epsilon$ such that the components of $C\verb|\|E_\epsilon$ all have diameter $\leq r$.

There is a logical problem with trying to establish the converse: we do not know how to even define the $\epsilon$-density of a subset if we do not first know that the graph is amenable.  (This is representative of a host of issues surrounding amenability, e.g. even in the simplest case $G\cong Z$ constructing the invariant measure requires the axiom of choice.)

We would like to propose a definition of a concept $k$-amenable which can be applied to a (single) noncompact metric space $X$ in the favorable case that they possess a transitive group of isometries.  This definition switches contexts: instead of discussing subsets $Y\subset X$ whose fractional volume $\text{vol }Y/\text{vol }X$ may make no sense, we discuss invariant measures on such subsets.  This approach was used by L.\ Bowen and C.\ Radin \cite{bowen} apparently following a suggestion of O.\ Schramm to create a context for studying (radius $=r$)-sphere packings in hyperbolic space.  Instead of asking: ``which packing is most dense,'' they ask ``which translation invariant measure on sphere packings assigns the highest probability to an arbitrary base point $*$ lying within one of the spheres''.  We conclude the appendix with our definition of $k$-amenable, intended as the non-compact or infinite measure counterpart to $k$-hyperfinite.  Just as $j$-hyperfinite implies $k$-hyperfinite for $j<k$, $j$-amenable implies $k$-amenable for $j<k$, and $0$-amenable should coincide with amenable in cases when both can be defined.

\noindent{\underline{Definition:}}
Let $X$ be a metric space with a transitive group $G$ of isometries.  $X$ is \emph{$k$-amenable} if $\forall$ $\epsilon>0$ $\exists$ $r>0$ and a $G$-invariant $\sigma$-additive (Borel) probability measure $\mu$ on closed subsets of $X$ so that:

\begin{enumerate}
    \item the probability of a base point $x\in X$ lying in a $\mu$ random subset is $>(1-\epsilon)$, and
    \item $\mu$ is supported on closed subsets $C_i$ admitting continuous maps $f_i:C_i\to K_i$, $K_i$ a $k$-complex, with $\text{diam}(f_i^{-1}(p))<r$ for all points $p\in K_i$.
\end{enumerate}


\begin{thebibliography}{99}
\bibitem{mtriverror} In Ref.~\onlinecite{mixedtriv}, states were considered on a two-dimensional lattice which were $(R,\epsilon)$ trivial for an $R$ growing logarithmically in system size, which does not quite meet this requirement.  However, Eq.~(8) in that paper contains an error, and should instead read $\l_\beta=R_{int} \sqrt{\exp((2R_{int}^2\beta) \log(V/\epsilon)}$ as can be seen by the requirement on the length scale immediately below the equation.  With this correction, the length scale indeed grows as the square-root of the log of the system size, allowing efficient evaluation of local observables.  It is also worth pointing out that what we consider to be trivial depends upon the application: for computational complexity reasons as considered here we are interested in efficient evaluation, while for problems in topological order we are often instead interested in whether the length $R$ scales linearly in system size or instead scales slower.
\end{thebibliography}
\end{document}